\documentclass[10pt,twocolumn,letterpaper]{article}

\usepackage[pagenumbers]{cvpr} % To force page numbers, e.g. for an arXiv version

\usepackage[dvipsnames]{xcolor}
\definecolor{runtimesecondcolor}{RGB}{200,200,255} % adjust RGB as needed
\definecolor{runtimesecondcolor}{RGB}{31,119,180} % Blue shade, adjust as needed
\newcommand{\TODO}[1]{\textbf{\color{red}[TODO: #1]}}
 \usepackage{array,multirow}
 \usepackage{amssymb}       % For \checkmark (built-in math checkmark)
\usepackage{pifont}        % For \ding symbols
\newcommand{\cmark}{\ding{51}}
\newcommand{\xmark}{\ding{55}}
\usepackage{tikz}
\usepackage{pgfplots}
\pgfplotsset{compat=1.18}

\usepackage{nicematrix}% NiceTabular and \rectanglecolor

\usepackage{xspace}
\usepackage{wrapfig}
\newcommand\ourmethod{AnyVIO\xspace}
\newcommand\ourmethodvo{AnyVO\xspace}

\newcommand\ourdataset{RaceDataset\xspace}
\newcommand\VGGTO{VGGT-$\Omega$\xspace}
\newcommand\mg[1]{{\color{blue}MG: #1}}
\newcommand\lc[1]{{\color{red}LC: #1}}
\newcommand\oa[1]{{\color{purple}OA: #1}}

\newcommand{\vect}[1]{\boldsymbol{#1}}
\newcommand{\mat}[1]{\boldsymbol{#1}}
\newcommand{\Real}{\mathbb{R}}

\newcommand{\Sigmoid}{\operatorname{sigmoid}}

\newcommand{\FrameIndex}{i}
\newcommand{\NextIndex}{j}
\newcommand{\WindowIndex}{k}

\newcommand{\SubmapSize}{N}

\newcommand{\Image}{\mat{I}}

\newcommand{\StepFunction}{f_{\mathrm{vis}}}
\newcommand{\StreamCache}{\mathsf{S}}

\newcommand{\CameraOutput}{\mathcal{C}}

\newcommand{\Quaternion}{\vect{q}}

\newcommand{\DenseOutput}{\mathcal{D}}

\newcommand{\MaskMap}{\mat{M}}

\newcommand{\VisualFeature}{\vect{f}^{\mathrm{V}}}

\newcommand{\State}{\vect{h}}

\usepackage{soul}
\setuldepth{foobar}

\providecommand{\tabheader}[1]{\textbf{#1}}

\newcommand\diminish[1]{#1}
\newcommand\bestcolor{white!80!black}
\newcommand\secondbestcolor{white!92!black}
\usepackage[export]{adjustbox}

\definecolor{cvprblue}{rgb}{0.21,0.49,0.74}
\usepackage[pagebackref,breaklinks,colorlinks,allcolors=cvprblue]{hyperref}

\title{AnyVIO: Feed-Forward Streaming Visual-Inertial Odometry Across Platforms}

\author{Omoruyi Atekha\\
MIT\\
{\tt\small oatekha@mit.edu}
\and
Marcus Greiff\\
TRI\\
{\tt\small marcus.greiff@tri.global}
\and
Ray Zhang\\
TRI\\
{\tt\small ray.zhang@tri.global}
\and
Luca Carlone\\
MIT\\
{\tt\small lcarlone@mit.edu}
}

\begin{document}
\maketitle
\begin{abstract}
Geometric foundation models (GFMs) estimate camera motion and scene structure from images without requiring expert parameter tuning, but either lack metric scale or rely on learned priors to infer scale, which is unobservable from monocular images alone. Inertial sensing offers a way to observe metric scale, but existing methods add inertial information through a filter or factor graph, which reintroduces bespoke parameter tuning and optimization costs that can grow with the trajectory.
To bridge this gap, we present \emph{\ourmethod}, a real-time, feed-forward streaming model for monocular visual-inertial odometry (VIO), which combines a streaming GFM with an inertial foundation model (IFM). Both the GFM and IFM predict the metric motion between consecutive frames, and a lightweight learned fusion head quantifies the reliability of each estimate and blends them in a single forward pass. Across nine datasets recorded from cars, legged and ground robots, and handheld and head-mounted devices, \ourmethod reduces the metric position error of its visual model by 36\%  while publishing each pose estimate 70\, ms after an image is received. 
We further deploy \ourmethod onboard two real vehicles at up to 140\, km/h, where it remains accurate in poor lighting and heavy dust.
\end{abstract}    

\begin{figure}[t]
    \centering
    \includegraphics[width=\columnwidth]{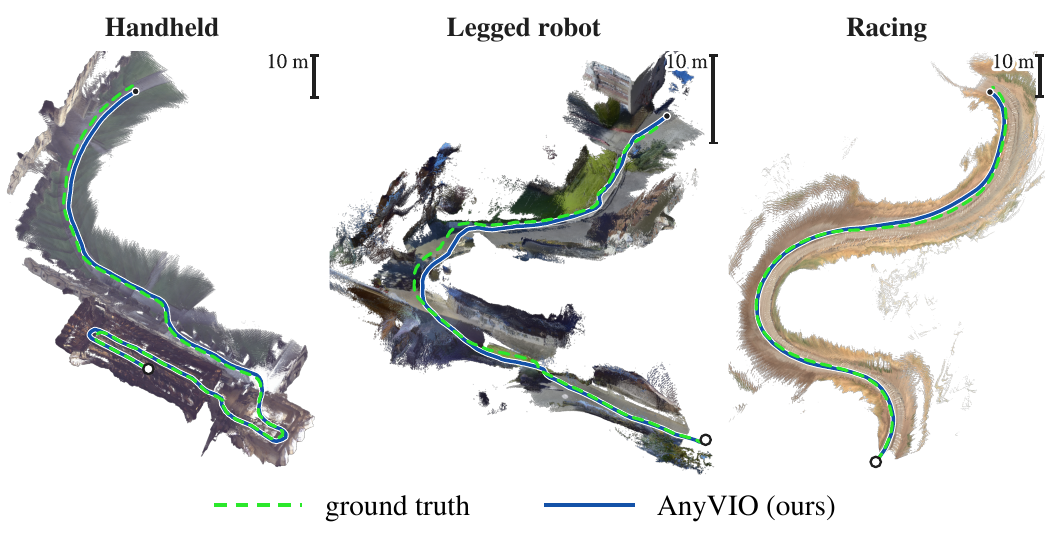}
    \vspace{-10pt}
    
    \caption{\textbf{\ourmethod}: a real-time, feed-forward model for monocular visual-inertial odometry (VIO), which combines a streaming GFM with an inertial foundation model to estimate metric odometry and 3D structure online across platforms and environments. \ourmethod's trajectory follows the ground truth closely. From left to right: Oxford Spires~\cite{tao2025spires} (handheld), GrandTour~\cite{frey2026grandtour} (legged robot), and our self-collected \ourdataset (vehicle racing).}
    \label{fig:teaser}
\end{figure}

\section{Introduction}\label{sec:intro}

Visual-Inertial Odometry (VIO) enables robots, autonomous vehicles, and virtual/augmented reality devices to estimate motion~\cite{li2013high, BloeschIROS15, bloesch2017iterated}. VIO extends visual odometry (VO) by incorporating high-frequency inertial measurements to improve localization and enable metric-scale monocular estimation \cite{VINS_source}.  Classical VIO methods~\cite{Forster2017manifold,VINS_source, VINS-Fusion_source, Rosinol20icra-Kimera, okvis2s_source, korovko2025cuvslam, ORBSLAM3_source} estimate camera poses by tracking image features across frames and fusing them with integrated IMU measurements, but require deployment-specific tuning. 
Early learned VIO methods~\cite{wei2025botvio,zhou2024dba,peng2024dvi,pan2024adaptive} encode both camera and IMU and regress fused motion without a filter or optimizer. However, they are trained on small visual backbones and typically constrained to one domain.

Geometric foundation models (GFMs) have been proposed as a more flexible alternative to classical structure-from-motion, SLAM, and VO  methods, where a single feed-forward model can be used to obtain odometry and 3D reconstruction without scene-dependent hyperparameter tuning. Rather than estimating poses through explicit feature tracking, GFMs~\cite{dust3r_cvpr24,leroy2024grounding,wang2025vggt,wang2026vggt,keetha2026mapanythinguniversalfeedforwardmetric,depthanything3} regress camera poses directly from pixels in one forward pass. \emph{Batch-wise} GFMs perform well in low-texture scenes where feature-based methods struggle, but their computational cost increases over time~\cite{rhee2026foundationvio}. \emph{Streaming} GFM variants~\cite{wang2025continuous, zhuo2026streaming, lan2025stream3r, chen2026lingbotmap, cheng2026horizonstream, cheng2026longstream, yuan2026infinitevggt} address this problem by processing one image at a time with bounded memory, but are currently limited to camera inputs alone and generally struggle to reliably estimate metric scale in different scenes~\cite{keetha2026mapanythinguniversalfeedforwardmetric, cheng2026horizonstream}. 

Recent efforts attempt to combine GFM with inertial measurements through preintegration and factor-graph optimizations~\cite{zhou2025mast3r,rhee2026foundationvio}. 
While carrying the promise of enabling or improving GFM metric-scale estimation, these methods 
reintroduce the need for expert parameter tuning and often lead to pipelines whose computational cost increases over time~\cite{rhee2026foundationvio}. On the other hand, 
inertial foundation models (IFMs)~\cite{zhao2025tartan} have recently been investigated 
as a way to obtain motion estimates using only IMU data, but still have the drawback of requiring platform-specific heads and being less accurate than classical VIO methods.

To enable cross-platform generalization, reduce the reliance on expert tuning, and leverage visual and inertial data, we introduce \emph{\ourmethod}, the first VIO streaming method to combine a 3D geometric foundation model (GFM) with an inertial foundation model (IFM). \ourmethod adapts a non-metric GFM to  incremental inference with bounded memory and augments it with a learned metric scale. A lightweight reliability-aware head fuses its motion estimate with that of an IFM without any filter, factor graph, or other optimization at inference. In cross-platform evaluations, a \emph{single \ourmethod model} demonstrates improved metric accuracy and faster frame-wise  time compared to state-of-the-art classical or GFM-based VO and  VIO methods on driving, indoor, and walking datasets.
We deploy \ourmethod on two vehicles driving at up to 140\,km/h and through heavy dust, where it remains robust to poor lighting and degraded sensing. 
In summary, 
the contributions of this work are:
\begin{itemize}
\item \textbf{Feed-forward Streaming VIO.} A streaming VIO method that combines a GFM with an IFM using reliability-aware learned fusion, inheriting the cross-domain generalization of the foundation models without filters, factor graphs, or optimization. The model will be released open-source upon acceptance.
\item \textbf{Cross-Domain Evaluation.} A comprehensive evaluation of \ourmethod, a single trained model, against state-of-the-art VO and VIO systems on nine datasets covering urban driving, racing, legged and ground robots, and hand-held devices. \ourmethod improves metric accuracy over its VO counterpart and outperforms the tested baselines despite using constant memory and compute. 
\item \textbf{Hardware Deployment.} In deployments on real vehicles, \ourmethod remains accurate and robust to poor lighting, reversed driving directions, and heavy dust. In each condition, it is more accurate with the IMU than without.
\end{itemize}

\section{Related Works}

\textbf{Optimization-Based VO and VIO.} Classical visual odometry and SLAM track image features and estimate motion in an optimization backend, as in ORB-SLAM3~\cite{ORBSLAM3_source} and cuVSLAM~\cite{korovko2025cuvslam}. Learned visual odometry maintains a similar backend to classical visual odometry and SLAM systems and learns a visual frontend: D3VO~\cite{yang2020d3vo} feeds learned depth, pose, and uncertainty to direct odometry, while DROID-SLAM~\cite{droidslam_source} and DPVO~\cite{dpvo_source} learn correspondences and refine poses with differentiable bundle adjustment. With a single camera, these methods lack metric scale. In parallel, visual-inertial odometry incorporates inertial measurements directly into the optimization via IMU preintegration~\cite{Forster2017manifold}, to improve pose estimates and recover metric scale, as in VINS-Fusion \cite{VINS-Fusion_source}, Kimera~\cite{Rosinol20icra-Kimera}, and OKVIS2~\cite{okvis2s_source}. Learned  variants~\cite{zhou2024dba, peng2024dvi, pan2024adaptive} add IMU factors to a  learned bundled adjustment. These systems rely on an optimizer at inference time and remain sensitive to degraded visual measurements  and tuned  sensor noise models.

\textbf{Geometric Foundation Models (GFMs).}
GFMs replace the feature-based frontend of classical VO with a single visual network trained on a large-scale, diverse set of 3D data. DUSt3R~\cite{dust3r_cvpr24} and MASt3R~\cite{leroy2024grounding} estimate dense pointmaps from an image pair and recover the relative camera pose from them, without feature tracking. Multi-view batch-wise methods~\cite{wang2025pi,keetha2026mapanythinguniversalfeedforwardmetric,depthanything3}, such as VGGT~\cite{wang_vggt_2025, wang2026vggtomega}, 
use a single transformer to attend a sequence of input images and predicts the geometry and trajectory of a short clip in one forward pass. These models are robust where classical VO feature tracking fails and need no-scene specific tuning. However, they  process a limited number of images at a time.

To adapt GFMs for longer input streams, one line of work makes GFMs themselves causal, processing each image once as it arrives.
Spann3R~\cite{wang20253d} and Point3R~\cite{wu2026point3r} keep an explicit spatial memory, CUT3R~\cite{wang2025continuous} and TTT3R~\cite{chen2026ttt3r} a recurrent state, and StreamVGGT~\cite{zhuo2026streaming} and STream3R~\cite{lan2025stream3r} a cache of past tokens read through causal attention. Recent models keep this cache bounded over long sequences by pruning, refreshing, or restricting it~\cite{yuan2026infinitevggt,cheng2026longstream,chen2026lingbotmap,cheng2026horizonstream,han2026revisiting, wang26stac}. Streaming GFMs need no optimizer, run in real time with bounded memory, and return each pose with low latency. One limitation is that they lack metric scale entirely or must predict it from appearance alone, and they cannot directly take advantage of IMU data even when available.

\textbf{GFMs in Factor Graph-based SLAMs.} GFMs can serve as frontends to be placed inside  classical optimization frameworks. MASt3R-SLAM~\cite{murai2025mast3r} builds a real-time dense SLAM system on MASt3R keyframes. VGGT-SLAM~\cite{maggio2025vggt,maggio2026vggt} runs VGGT on short submaps and aligns them in a factor graph with loop closures. VGGT-Motion~\cite{xiong2026vggt} aligns submaps with similarity transforms, and UniSim-SLAM~\cite{lee2026unisim} jointly optimizes keyframe and submap poses in a SIM(3) factor graph. ScaRF-SLAM~\cite{zhang2026scarf} instead tracks the camera with classical visual SLAM and uses the GFM only to build the map. While these systems make submaps consistent with one another, they can at best regress metric scale from images; MASt3R-Fusion~\cite{zhou2025mast3r} and FoundationVIO~\cite{rhee2026foundationvio} estimate scale  by adding IMU preintegration to the factor graph, and VI3~\cite{lozano2026vi3groundingpretrained3d} encodes a preintegrated IMU trajectory into the GFM as a pose input for short clips. These methods integrate the IMU and optimize at inference, but the complexity may grow with sequence length (\cref{tab:runtime:scaling}).

\textbf{VIO with Learned Inertial Odometry.}
Learned inertial odometry replaces analytic preintegration with a network that regresses velocity or displacement from a short window of IMU measurements~\cite{9134860,qiu2025airio,qiu2026ssio}, often fused afterwards in a filter. 
TartanIMU~\cite{zhao2025tartan}  pretrains a shared backbone with platform-specific heads across cars, legged robots, drones, and pedestrians, as an inertial foundation model.
End-to-end learned VIO methods, from VINet~\cite{Clark_Wang_Wen_Markham_Trigoni_2017} to recent transformer-based models~\cite{kurt2024vift,wei2025botvio},  fuse visual and inertial features directly, but typically rely on compact encoders trained on individual benchmarks.  \ourmethod instead combines a streaming GFM with an inertial foundation model through a learned, reliability-weighted fusion head, benefiting from large-scale visual and inertial pretraining without inference-time optimization.

\begin{figure*}[t]
  \centering
  \includegraphics[width=\textwidth]{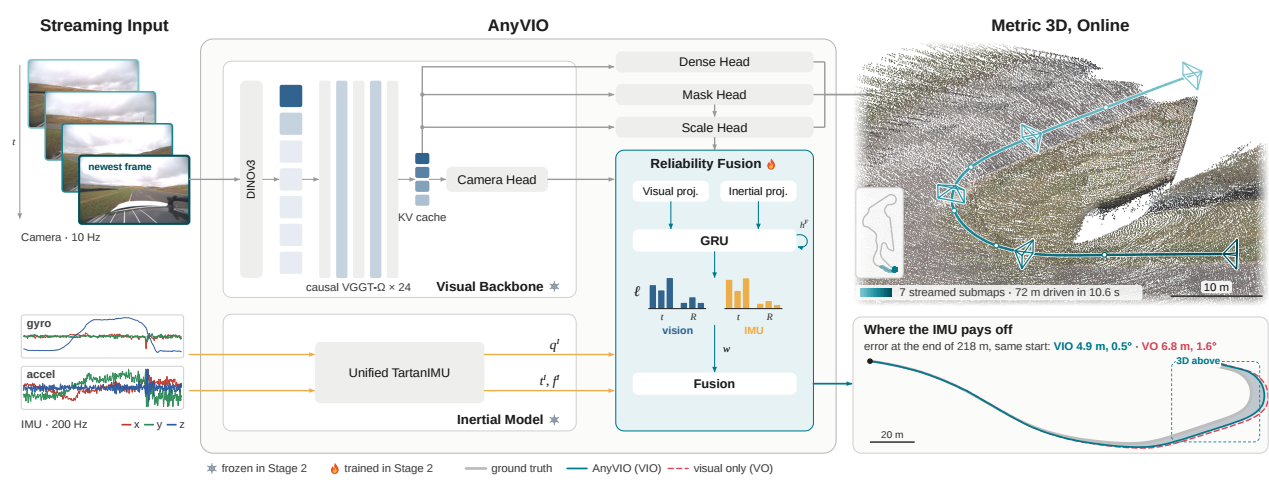}
  \vspace{-15pt}
  \caption{\textbf{\ourmethod overview}.
  Images are processed one at a time by a causal \VGGTO backbone that predicts camera motion, metric scale, dense depth, and a non-ambiguity mask. The inertial model integrates the gyroscope and predicts velocity from IMU histories. The reliability fusion head weighs both incremental motion estimates and outputs metric poses and a dense 3D map online. Fusion reduces drift over a 218\,m-long trajectory from 10.4\,m (VO) to 5.1\,m (VIO).}
  \label{fig:method:overview}
\end{figure*}

\section{Method}\label{sec:method}
\ourmethod has three main components outlined in~\cref{fig:method:overview}. A streaming visual backbone (\cref{sec:method:streamingvisual}) processes one image at a time and predicts visual features and metric camera motion. An IFM (\cref{sec:method:inertial}) produces inertial features and an independent odometry estimate. Finally, a fusion module (\cref{sec:method:inertial}) predicts the reliability of both estimates and blends them into the output of \ourmethod. %The foundation models are frozen while the fusion head is trained. %Furthermore, we also train a batch variant, in which the backbone attends bidirectionally within each submap; it gives the batch versions of \ourmethodvo and \ourmethod reported in our experiments.
We denote the variant of \ourmethod that only uses the  visual pathway as \ourmethodvo.

\subsection{\ourmethodvo: Visual Foundation Model}\label{sec:method:streamingvisual}

Inspired by \cite{vggtslam_source, maggio2026vggt}, we process fixed \emph{submaps} that share one frame, reset the cache at each boundary, and chain the submaps into a trajectory similar to~\cite{leroy2024grounding}, producing a bounded streaming window. We develop our streaming visual backbone from \VGGTO~\cite{wang2026vggt}, which has recently shown promise in VIO settings~\cite{rhee2026foundationvio, lozano2026vi3groundingpretrained3d} and is generally superior to VGGT~\cite{wang2025vggt} on which~\cite{zhuo2026streaming,rhee2026foundationvio, lozano2026vi3groundingpretrained3d} are built.

%\VGGTO~\cite{wang2026vggt} processes a batch of $\SubmapSize$ images using a DINOv3 encoder~\cite{simeoni2025dinov3}, bidirectional inter-frame attention, and decodes camera parameters, depth, and confidence in scene-normalized units in a single pass, referred to as a \emph{submap}. It does not directly provide incremental metric odometry. We retain its image encoder, alternating-attention aggregator, camera head, and dense decoder, and introduce frame-causal execution, a non-ambiguity mask prediction, and a separate metric-scale head. The streaming visual backbone processes images incrementally and predicts a metric scale for every frame from the frames received. As in~\cite{zhuo2026streaming,lan2025stream3r}, causality is obtained by restricting inter-frame attention to past frames. Unlike~\cite{zhuo2026streaming}, we fine-tune the causal model directly instead of distilling a bidirectional teacher.

%metric scale is finalized once a submap of $\SubmapSize$ frames is complete.

\noindent\textbf{Streaming Visual Backbone.}
%\lc{the notation is a bit clunky - do we really need to distinguish local from global frame indices?}
%\lc{is the downsampling applied to each submap? maybe we should rephrase saying that we select keyframes every k frames and then rephrase the discussion below in terms of keyframes?}
%We index images globally after subsampling. Submap $\WindowIndex$ is constructed from a set of $\SubmapSize$ images with a first image index $\SubmapStart_{\WindowIndex} = 1+\WindowIndex(\SubmapSize-1)$. 
Let $\Image_{\WindowIndex,\FrameIndex}\in\Real^{W\times H\times 3}$ denote the $i$th keyframe (RGB image) used in constructing the $k$th submap, where $i=1,...,N$.
%Its local image index is $\FrameIndex\in[\SubmapSize]$. 
For each keyframe, let 
$
\CameraOutput_{\WindowIndex,\FrameIndex}
%=(\Pose_{\WindowIndex,\FrameIndex},\FovParameters_{\WindowIndex,\FrameIndex})
$ 
contain the decoded camera-to-world pose in the submap frame and field-of-view parameters. Let $\DenseOutput_{\WindowIndex,\FrameIndex}
%=(\DepthMap_{\WindowIndex,\FrameIndex},\ConfidenceMap_{\WindowIndex,\FrameIndex})
$ be the dense depth and confidence maps as defined in \VGGTO. We add a non-ambiguity mask and scale head, predicting $\MaskMap_{k,i}\in[0,255]^{W\times H}$ and $s_{k,i}>0$, respectively. The streaming visual backbone $\StepFunction$ is then
\begin{equation}\label{eq:IO}
\hspace{-2pt}(\CameraOutput_{\WindowIndex,\FrameIndex}, \DenseOutput_{\WindowIndex,\FrameIndex}, \MaskMap_{\WindowIndex,\FrameIndex}, 
s_{\WindowIndex,\FrameIndex},
\StreamCache_{\WindowIndex,\FrameIndex}) \hspace{-2pt}=\hspace{-2pt}
\StepFunction(\Image_{\WindowIndex,\FrameIndex}, \StreamCache_{\WindowIndex,\FrameIndex-1}),\hspace{-2pt}
\end{equation}
where $\StreamCache$ denotes a streaming KV-cache. The metric camera odometry is computed and externalized on each call of~\eqref{eq:IO}.
%From these outputs and inetermediary latents, we accumulate visual and geometric descriptors, $\VisualFeature_{\WindowIndex,\NextIndex}$ and $\GeometricFeature_{\WindowIndex,\NextIndex}$, respectively. For every frame $i > 1$, we predict the metric scale of the edge from frame $i - 1$ to $i$ from the frames received so far, 
%
%\begin{equation}
%    s_{\WindowIndex,\FrameIndex} = f_\text{scale}\Big(\big(\VisualFeature_{\WindowIndex,j},\,\vect{f}^G_{\WindowIndex,j}\big)_{j=1}^{\FrameIndex}\Big), \qquad \FrameIndex = 2,\dots,\SubmapSize,
%\end{equation}
%
%and chain the scaled edges into metric camera poses, so that every frame receives a metric pose when it arrives. The first frame of each submap is shared with the previous one and keeps its pose.

Similar to~\cite{zhuo2026streaming,lan2025stream3r,cheng2026longstream}, inter-frame attention at frame $i$ is restricted to tokens from frames $\NextIndex\leq\FrameIndex$. We apply this to the camera head and the aggregator module of \VGGTO, making it causal. 
%Frame attention and the dense heads act on single frames and need no restriction. 
Each new image passes through the image encoder and aggregator once, except for a shared boundary image, which is re-encoded when a new submap begins. %Queries attend to previously cached and newly computed keys and values
%\begin{equation}
%    \operatorname{Attn}_{\FrameIndex}
%    =
%    \Softmax\Bigg(
%        \frac{
%            \Query_{\FrameIndex}
%            \Key_{\leq\FrameIndex}^{\Transpose}
%        }{\sqrt{\HeadWidth}}
%    \Bigg)
%    \Value_{\leq\FrameIndex}.
%\end{equation}

The cache $\StreamCache_{k,i-1}$ in~\eqref{eq:IO} contains the keys and values for inter-frame attention layers of frames processed prior to $\Image_{k,i}$. %It retains the frame counter, the reusable spatial rotary-position embeddings, over which the camera head is reevaluated, and the recurrent state of the scale head. 
At each submap boundary ($i=N$), the cache and local buffers are reset, analogous to the KV-cache logic in~\cite{cheng2026longstream}.

\noindent\textbf{Mask Head.}
Inspired by MapAnything~\cite{keetha2026mapanythinguniversalfeedforwardmetric}, we add a non-ambiguity mask head that marks pixels with well-defined geometry. In our driving experiments, it is supervised to exclude the sky and the vehicle body. The fraction of unambiguous pixels is informative in assessing reliability and used as an input to the scale head.

\noindent\textbf{Scale Head.} \ourmethodvo predicts poses in scene-normalized units, so each submap is estimated at its own normalized scale. %To chain frames into a trajectory, and to fuse that trajectory with inertial measurements, every edge must be expressed at metric scale. 
As each frame arrives, the scale head predicts metric scale from a visual feature, $\VisualFeature$, which contains the camera-head latent, pixel-wise statistics of the predicted depth confidence and the mask coverage, and the camera pose relative to the first frame of the submap. A GRU accumulates these inputs continuously, and an MLP on its state predicts the log-scale of an incoming frame $\lambda_k$. The state of the GRU is reset each time the cache is reset ($i=N$). Additional details on the training of the mask and scale head, and GRU initialization and states are given in the supplementary.

\subsection{Inertial Foundation Model}\label{sec:method:inertial}
To generate IMU features and odometry that generalize across various datasets, we consider the recent TartanIMU~\cite{zhao2025tartan}. This IFM combines a shared ResNet--LSTM encoder with separate velocity heads for each platform (car, quadruped, drone, and human), as their speed profiles differ substantially. We adapt this to be a single shared velocity head, so that one model serves multiple platforms. To facilitate this, we fine-tune with platform-balanced sampling and a normalized smooth L1~\cite{10.1109/ICCV.2015.169} loss on the velocities
\begin{equation}\label{eq:imu_velocity_loss}
\mathcal L_{\mathrm{IMU}}
=
\frac{1}{N}\sum_n
\operatorname{SmoothL1}_{\beta=1}\!\left(
\frac{\hat{\boldsymbol v}_n^B-\boldsymbol v_n^{B,*}}{c_n}
\right).
\end{equation}
For a given sample $n$, $\hat{\boldsymbol v}_n^B$ and ${\boldsymbol v}_n^{B}$ are the predicted and ground-truth body velocities. For each sample, the scalar $c_n$ normalizes the error by a scale derived from the standard deviation of the training velocities in its dataset, shared across all three velocity components (see the supplementary material for details). The output of the fine-tuned IFM is a translational velocity in the body frame, $\hat{\boldsymbol v}^B$, and a feature vector, $\boldsymbol{f}^I$, which are used as inputs to the fusion module. The former is integrated along with the gyroscopic measurements to produce a relative pose between consecutive key-frames, and rotated into a camera frame using known IMU-camera extrinsics.

\subsection{\ourmethod: VIO Fusion}\label{sec:method:fusion}

\noindent\textbf{Motion Hypotheses.} From frame $i$ to $i+1$, the visual backbone and the inertial model estimate the metric translation and rotation quaternion as $(\vect{t}_i^V,\Quaternion_i^V)$ and $(\vect{t}_i^I,\Quaternion_i^I)$, respectively. The fusion head uses a geometric descriptor and the VIO embedding to predict the reliability of the two motion hypotheses before fusing them. The geometric descriptor, $\vect{g}_i$, contains the motion estimates, the difference between them in translation and rotation, statistics of the raw IMU signal over the time window of the motion, and measures of how well the two estimates agree over the time window. 
%This provides a comprehensive sense of how the GFM and IFM disagree.
The VIO embedding,  $\vect{z}_i\in\Real^{64}$, is a compact projection of the visual and inertial features $\VisualFeature$ and $\vect{f}^I$ of two consecutive frames, capturing both the inertial and visual information. Full details are reported in the supplementary.

We introduce a simple model to fuse this information: a GRU with two output heads, a correction head and a reliability head. The correction head, $f_c$, predicts two varying corrections to the inertial hypothesis: a bounded inertial log-scale, $\gamma_I$, and gyroscope bias, ${\vect{b}}_i$. This accounts for slowly varying biases and errors in the inertial estimate. The reliability head, $f_\ell$, uses the state $\State_i^F$ from a GRU network along with $\vect{z}_i$ to predict
$\vect{\ell}_i \in \Real^{12}$, the log error variances for the visual and inertial translation and rotation, respectively. This is intended to capture sensor changes with environmental and dynamic conditions. We also include an MLP, $f_{\rho}$, which computes a correction $\rho_V$ to the scale of the visual translation, predicted from the visual features alone:
\begin{equation}\label{eq:fusion_predictions}
\begin{aligned}
\State_i^F&=\operatorname{GRU}(\vect{g}_i,\State_{i-1}^F), &
(\gamma_i,{\vect{b}}_i)&=f_c(\State_i^F),\\
\vect{\ell}_i&=f_\ell(\vect{z}_i,\State_i^F), &
\rho_V&=f_\rho(\vect{z}_i).
\end{aligned}
\end{equation}
These corrections $\gamma_i$ and $\vect{b}_i$ are applied to the inertial hypothesis, as $\hat{\vect{t}}_i^I = e^{\gamma_i}\vect{t}_i^I$ and $\hat{\Quaternion}_i^I = \Quaternion_i^I \otimes \operatorname{Exp}(-\vect{b}_i\Delta t_i)$, where $\Delta t_i$ is the time step between the two camera frames.

\noindent\textbf{Motion Update.} Rotations and translations from the GFM and IFM are combined separately. Over the interval between $i$ and $i+1$, the gyroscope, $\hat{\Quaternion}_i^I$ corrects the visual rotation, $\Quaternion_i^V$,  to get $\Quaternion_i^F$. To keep the translation fusion independent of rotation, we assume both translation hypotheses share the same direction and differ only in scale. Over the $\SubmapSize$ most recent frames, we sum the length of the submap to step $i$, giving the distances $L^V$ and $L^I$. Their ratio is the inertial scale $\rho_I=L^I/L^V$, which describes how much longer or shorter the IFM estimated the translation compared to the GFM. The inertial translation hypothesis is $\rho_I\,\vect{t}_i^V$, and the visual translation hypothesis is $\rho_V\,\vect{t}_i^V$. For each coordinate of the translation and rotation, the two estimates are weighed by the difference, $d$, between the visual and inertial log-variance, $d^u_{i,a}=\ell^{V,u}_{i,a}-\ell^{I,u}_{i,a}$ for $u\in\{t,r\}$ and coordinate $a$. The inertial weight is $\vect{w}_i^u$ with elements
\begin{equation}\label{eq:fusion_weights}
w^u_{i,a}=\Sigmoid(d^u_{i,a})\Sigmoid(\tfrac12 (d^u_{i,a}-1)).
\end{equation}
The fused motion is then computed as
\begin{equation}\label{eq:fusion_update}
\begin{aligned}
\vect{t}_i^F&=\big(\rho_V+\vect{w}_i^t(\rho_I-\rho_V)\big)\odot\vect{t}_i^V,\\
\Quaternion_i^F&=\Quaternion_i^V\otimes\operatorname{Exp}\!\Big(\vect{w}_i^r\odot\operatorname{Log}\big((\Quaternion_i^V)^{-1}\otimes\hat{\Quaternion}_i^I\big)\Big),
\end{aligned}
\end{equation}
where $\odot$ denotes coordinate-wise multiplication. Rotation is then fused with the bias-corrected gyroscope, while the IMU affects translation only through its scale.

\noindent\textbf{Fusion Loss and Training.}
The VIO fusion head is trained after training the IFM (\cref{sec:method:inertial}) and streaming GFM (\cref{sec:method:streamingvisual}). Training it jointly with the visual pose and scale layers gave lower validation accuracy (see supplementary), so the fusion head was trained with both models frozen. The fusion head is trained with the following loss on sequences of consecutive windows, so that the GRU state and the smoothed corrections are carried from one window to the next as they would during inference:
\begin{equation}\label{eq:fusion_loss}
    \mathcal{L}_F = \mathcal{L}_\text{pose} + \mathcal{L}_\text{rel} + \mathcal{L}_\text{corr}.
\end{equation}

$\mathcal{L}_\text{pose}$ is a Huber loss on the error of the fused motion~\eqref{eq:fusion_update} with respect to the ground truth. $\mathcal{L}_\text{rel}$, the Gaussian negative log-likelihood of each hypothesis's error under its predicted variance, trains the reliability head. The $\mathcal{L}_\text{corr}$, supervises the corrections $\gamma_i$ and $\vect{b}_i$ with targets computed from ground truth. At each training step, the same number of sequences is sampled from each dataset, so that the model is trained evenly across platforms. Loss hyperparameters and training details are provided in the supplementary material.
\begin{figure}[t]
    \centering

    % Mosaic dimensions:
    \def\ncols{6}
    \def\nrows{4}

    % Tile dimensions in cm:
    \def\tilewidth{1.3}
    \def\tileheight{0.6}

    \begin{tikzpicture}[
        every node/.style={
            anchor=south west,
            inner sep=0pt,
            outer sep=0pt
        }
    ]
        % Six filename prefixes:
        % Top:    tarmac Train, In-distribution, OOD
        % Bottom: offroad Train, In-distribution, OOD
        \foreach \name [count=\groupindex from 0] in {
            leia_train,
            leia_id,
            leia_ood,
            odin_train,
            odin_id,
            odin_ood%
        } {
            \pgfmathtruncatemacro{\groupcol}{mod(\groupindex,3)}
            \pgfmathtruncatemacro{\grouprow}{1-floor(\groupindex/3)}

            \foreach \imageindex in {1,...,4} {
                \pgfmathtruncatemacro{\localcol}{mod(\imageindex-1,2)}
                \pgfmathtruncatemacro{\localrow}{1-floor((\imageindex-1)/2)}

                \node at (
                    {(2*\groupcol+\localcol)*\tilewidth},
                    {(2*\grouprow+\localrow)*\tileheight}
                ) {%
                    \includegraphics[
                        width=\tilewidth cm,
                        height=\tileheight cm
                    ]{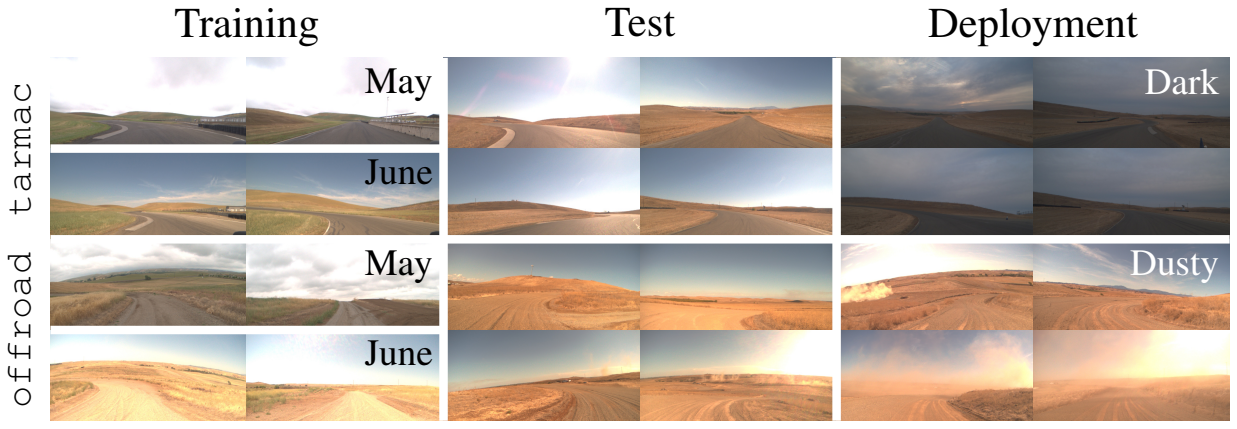}%
                };
            }
        }

        \node[anchor=north]
            at ({1*\tilewidth},{4.55*\tileheight}) {\footnotesize Training};
        \node[anchor=north]
            at ({3*\tilewidth},{4.55*\tileheight}) {\footnotesize Test};
        \node[anchor=north]
            at ({5*\tilewidth},{4.55*\tileheight}) {\footnotesize Deployment};

        \iffalse
        \node[anchor=north east, outer sep=2pt]
            at ({6*\tilewidth},{4*\tileheight}) {{\color{white}\scriptsize exp 216}};
        \node[anchor=north east, outer sep=2pt]
            at ({6*\tilewidth},{2*\tileheight}) {{\color{white}\scriptsize exp 072}};
        \node[anchor=north east, outer sep=2pt]
            at ({4*\tilewidth},{2*\tileheight}) {{\color{white}\scriptsize exp 069}};
        \node[anchor=north east, outer sep=2pt]
            at ({4*\tilewidth},{4*\tileheight}) {{\color{white}\scriptsize exp 179}};
        \fi

        \node[anchor=north east, outer sep=2pt]
            at ({6*\tilewidth},{4*\tileheight}) {{\color{white}\scriptsize Dark}};
        \node[anchor=north east, outer sep=2pt]
            at ({6*\tilewidth},{2*\tileheight}) {{\color{white}\scriptsize Dusty}};

        \node[anchor=north east, outer sep=2pt]
            at ({2*\tilewidth},{4*\tileheight}) {{\color{black}\scriptsize May}};
        \node[anchor=north east, outer sep=2pt]
            at ({2*\tilewidth},{3*\tileheight}) {{\color{black}\scriptsize June}};
        \node[anchor=north east, outer sep=2pt]
            at ({2*\tilewidth},{2*\tileheight}) {{\color{black}\scriptsize May}};
        \node[anchor=north east, outer sep=2pt]
            at ({2*\tilewidth},{1*\tileheight}) {{\color{black}\scriptsize June}};

        \node[anchor=south, rotate=90]
            at ({-0.1*\tilewidth},{3*\tileheight}) {\scriptsize\texttt{tarmac}};
        \node[anchor=south, rotate=90]
            at ({-0.1*\tilewidth},{1*\tileheight}) {\scriptsize\texttt{offroad}};

        \draw[white, ultra thick]
            ({0*\tilewidth},{2*\tileheight}) --
            ({6*\tilewidth},{2*\tileheight});
        \draw[white, ultra thick]
            ({2*\tilewidth},{0*\tileheight}) --
            ({2*\tilewidth},{4*\tileheight});
        \draw[white, ultra thick]
            ({4*\tilewidth},{0*\tileheight}) --
            ({4*\tilewidth},{4*\tileheight});
        \draw[white, ultra thick]
            ({0*\tilewidth},{1*\tileheight}) --
            ({2*\tilewidth},{1*\tileheight});
        \draw[white, ultra thick]
            ({0*\tilewidth},{3*\tileheight}) --
            ({2*\tilewidth},{3*\tileheight});
    \end{tikzpicture}
    \vspace{-7pt}

    \caption{\textbf{The \ourdataset.} Sample images from training (left) and test sequences (center), images from challenging deployments with poor lighting and visibility (right).}
    \label{fig:mosaic}
    \vspace{-12pt}
\end{figure}

\section{Experimental Results}
 We train and evaluate AnyVIO on a collection of visual-inertial datasets across different settings (\cref{tab:datasets}): car racing (on- and off-road), urban driving, legged and ground robots, and handheld and head-mounted devices.
%\subsection{Overview}
%\label{sec:data}
 Subsets of the datasets support different training stages: the inertial foundation model (I), the visual foundation model (V), the fusion head (F), or for evaluation (E). KITTI Odometry \cite{geiger2013vision} \cite{Geiger2012CVPR}, ScaRF \cite{zhang2026scarf}, and OpenLORIS \cite{shi2019openlorisscene} are held-out for evaluation only and excluded from the training dataset, marked with $\ddagger$ in our tables. For IMU training, we include the original data from TartanIMU~\cite{zhao2025tartan}. %comprising including its four modalities Car, Drone, Quadruped, and Human.
We further collect a self-recorded vehicle racing dataset, \emph{\ourdataset}, as shown in  Fig.~\ref{fig:mosaic}, to  evaluate extreme motions and dusty, repetitive, and glary camera observations, see the supplementary for further details. Lastly, we deploy \ourmethod on the real hardware in the nominal and out-of-distribution racing scenes.

\begin{table}[t!]
\centering\small
\captionsetup{font=small}
\setlength{\tabcolsep}{5pt}
\renewcommand{\arraystretch}{1.0}
\caption{\textbf{Dataset inventory and usage.} \tabheader{I}: used for IFM training; \tabheader{V}: used for GFM training; \tabheader{F}: used for fusion-head training; \tabheader{E}: used for evaluation (Secs.~\ref{sec:eval:driving}--\ref{sec:eval:indoor}). 
%(Secs.~\ref{sec:eval:driving}--\ref{sec:eval:indoor}). %Sequences are distinct recordings in our selected training, validation, and test pools, not full-release sizes; training counts precede visual depth filtering. A test recording may contain multiple independently evaluated segments. Rates are median prepared frequencies (Hz). --: unused or unavailable.
}
\label{tab:datasets}

\vspace{-8pt}
\resizebox{\columnwidth}{!}{%
\begin{tabular}{@{}l@{\hspace{9pt}}rr@{\hspace{9pt}}rrr@{\hspace{9pt}}cccc@{}}
\toprule
\multirow{2}{*}{\tabheader{Dataset}} & \multicolumn{2}{c}{\tabheader{Rates (Hz)}} & \multicolumn{3}{c}{\tabheader{Sequences}} & \multicolumn{4}{c}{\tabheader{Usage}} \\
\cmidrule(lr){2-3}\cmidrule(lr){4-6}\cmidrule(lr){7-10}
 & \tabheader{Cam} & \tabheader{IMU} & \tabheader{Train} & \tabheader{Val} & \tabheader{Test} & \tabheader{I} & \tabheader{V} & \tabheader{F} & \tabheader{E} \\
\midrule
\ourdataset & 29 & 501 & 40 & 18 & 6 & $\checkmark$ & $\checkmark$ & $\checkmark$ & $\checkmark$ \\
KITTI-360 \cite{Liao2022PAMI} & 10 & 100 & 4 & 2 & 2 & $\checkmark$ & $\checkmark$ & $\checkmark$ & $\checkmark$ \\
KITTI Odometry$^\ddagger$ \cite{geiger2013vision} \cite{Geiger2012CVPR} & 10 & 100 & -- & -- & 10 & -- & -- & -- & $\checkmark$ \\
TartanDrive 2 \cite{sivaprakasam2024tartandrive} & 10 & 100 & 14 & 4 & 4 & $\checkmark$ & $\checkmark$ & $\checkmark$ & $\checkmark$ \\
nuScenes \cite{DBLP:journals/corr/abs-1903-11027}& 10 & 100 & 60 & 2 & -- & -- & $\checkmark$ & $\checkmark$ & -- \\
\midrule
GrandTour \cite{frey2026grandtour} & 10 & 201 & 11 & 8 & 8 & $\checkmark$ & $\checkmark$ & $\checkmark$ & $\checkmark$ \\
Oxford Spires \cite{tao2025spires} & 20 & 400 & 7 & 3 & 3 & $\checkmark$ & $\checkmark$ & $\checkmark$ & $\checkmark$ \\
Aria Digital Twin \cite{10376711} & 10 & 1002 & 32 & 8 & 16 & $\checkmark$ & -- & $\checkmark$ & $\checkmark$ \\
Nymeria \cite{nymeria24} \cite{nymeriaplus26} & 15 & 1001 & 13 & 3 & -- & $\checkmark$ & $\checkmark$ & $\checkmark$ & -- \\
M2DGR \cite{yin2021m2dgr}& 15 & 146 & 7 & 2 & -- & $\checkmark$ & $\checkmark$ & $\checkmark$ & -- \\
ScaRF$^\ddagger$ \cite{zhang2026scarf} & 30 & 1000 & -- & -- & 3 & -- & -- & -- & $\checkmark$ \\
OpenLORIS$^\ddagger$ \cite{shi2019openlorisscene} & -- & -- & -- & -- & 5 & -- & -- & -- & $\checkmark$ \\
\midrule
TartanIMU \cite{zhao2025tartan} $^{\dagger}$ & -- & 200 & 395 & 80 & -- & $\checkmark$ & -- & -- & -- \\
\bottomrule
\end{tabular}}
%\par\vspace{2pt}{\raggedright\footnotesize $\dagger$: all 395 training recordings; 80 validation recordings remain separate.\par}
\end{table}

%\section{Results}

% Main driving table RPE
\newcommand\tmpnum{\color{red}todo}

\begin{table}[t!]
\caption{\textbf{Performance on driving datasets.} RPE per 100 m with a monocular camera and loop closures disabled. \tabheader{Opt}: iterative optimization at inference. \tabheader{IMU}: IMU enabled. Mean relative errors are reported: translation in \% and rotation in $^\circ$/100 m. Completion counts are in the corresponding APE table. 
Best and second-best results among fully completed evaluations are highlighted in gray, separately within each VO/VIO and streaming/batched block.}
\vspace{-7pt}
\label{tab:main:rpe}
\centering
\resizebox{\columnwidth}{!}{%
\renewcommand{\arraystretch}{1.08}
\setlength{\tabcolsep}{3pt}
\begin{NiceTabular}{@{\hspace{2pt}}c@{\hspace{6pt}}l|cc|cc|cc|cc|cc}[code-before={%
\rectanglecolor{\secondbestcolor}{5-5}{5-5}
\rectanglecolor{\secondbestcolor}{5-7}{5-7}
\rectanglecolor{\secondbestcolor}{5-9}{5-9}
\rectanglecolor{\secondbestcolor}{8-12}{8-12}
\rectanglecolor{\secondbestcolor}{8-6}{8-6}
\rectanglecolor{\secondbestcolor}{8-8}{8-8}
\rectanglecolor{\secondbestcolor}{8-10}{8-10}
\rectanglecolor{\secondbestcolor}{8-11}{8-11}
\rectanglecolor{\bestcolor}{9-5}{9-5}
\rectanglecolor{\bestcolor}{9-6}{9-6}
\rectanglecolor{\bestcolor}{9-7}{9-7}
\rectanglecolor{\bestcolor}{9-8}{9-8}
\rectanglecolor{\bestcolor}{9-9}{9-9}
\rectanglecolor{\bestcolor}{9-10}{9-10}
\rectanglecolor{\bestcolor}{9-11}{9-11}
\rectanglecolor{\bestcolor}{8-12}{8-12}
\rectanglecolor{\bestcolor}{4-12}{4-12}
%%%
\rectanglecolor{\secondbestcolor}{11-5}{11-5}
\rectanglecolor{\secondbestcolor}{11-6}{11-6}
\rectanglecolor{\secondbestcolor}{11-7}{11-7}
\rectanglecolor{\secondbestcolor}{11-8}{11-8}
\rectanglecolor{\secondbestcolor}{11-9}{11-9}
\rectanglecolor{\secondbestcolor}{11-10}{11-10}
\rectanglecolor{\secondbestcolor}{11-11}{11-11}
\rectanglecolor{\secondbestcolor}{11-12}{11-12}
\rectanglecolor{\bestcolor}{12-5}{12-5}
\rectanglecolor{\bestcolor}{12-6}{12-6}
\rectanglecolor{\bestcolor}{12-7}{12-7}
\rectanglecolor{\bestcolor}{12-8}{12-8}
\rectanglecolor{\bestcolor}{12-9}{12-9}
\rectanglecolor{\bestcolor}{12-10}{12-10}
\rectanglecolor{\bestcolor}{12-11}{12-11}
\rectanglecolor{\bestcolor}{12-12}{12-12}
%%%
\rectanglecolor{\secondbestcolor}{13-5}{13-5}
\rectanglecolor{\secondbestcolor}{13-6}{13-6}
\rectanglecolor{\secondbestcolor}{13-7}{13-7}
\rectanglecolor{\secondbestcolor}{13-9}{13-9}
\rectanglecolor{\secondbestcolor}{13-12}{13-12}
\rectanglecolor{\secondbestcolor}{15-8}{15-8}
\rectanglecolor{\secondbestcolor}{15-10}{15-10}
\rectanglecolor{\secondbestcolor}{15-11}{15-11}
\rectanglecolor{\bestcolor}{16-5}{16-5}
\rectanglecolor{\bestcolor}{16-6}{16-6}
\rectanglecolor{\bestcolor}{16-7}{16-7}
\rectanglecolor{\bestcolor}{16-8}{16-8}
\rectanglecolor{\bestcolor}{16-9}{16-9}
\rectanglecolor{\bestcolor}{16-10}{16-10}
\rectanglecolor{\bestcolor}{16-11}{16-11}
\rectanglecolor{\bestcolor}{16-12}{16-12}
\rectanglecolor{\bestcolor}{16-12}{16-12}
%%%
\rectanglecolor{\secondbestcolor}{17-7}{17-7}
\rectanglecolor{\secondbestcolor}{17-8}{17-8}
\rectanglecolor{\secondbestcolor}{17-12}{17-12}
\rectanglecolor{\secondbestcolor}{18-5}{18-5}
\rectanglecolor{\secondbestcolor}{18-6}{18-6}
\rectanglecolor{\bestcolor}{18-7}{18-7}
\rectanglecolor{\bestcolor}{18-8}{18-8}
\rectanglecolor{\bestcolor}{18-9}{18-9}
\rectanglecolor{\bestcolor}{18-10}{18-10}
\rectanglecolor{\secondbestcolor}{18-11}{18-11}
\rectanglecolor{\bestcolor}{18-12}{18-12}
\rectanglecolor{\bestcolor}{19-5}{19-5}
\rectanglecolor{\bestcolor}{19-6}{19-6}
\rectanglecolor{\secondbestcolor}{19-9}{19-9}
\rectanglecolor{\secondbestcolor}{19-10}{19-10}
\rectanglecolor{\bestcolor}{17-11}{17-11}
}]
\toprule
\multicolumn{2}{c|}{\tabheader{Method}} & \tabheader{Opt.} & \tabheader{IMU} & \multicolumn{2}{c|}{\tabheader{\ourdataset}} & \multicolumn{2}{c|}{\tabheader{KITTI-360}} & \multicolumn{2}{c|}{\tabheader{TartanDrive 2}} & \multicolumn{2}{c}{\tabheader{KITTI Odom.$^\ddagger$}} \\
 &  &  &  & \tabheader{Trans.} & \tabheader{Rot.} & \tabheader{Trans.} & \tabheader{Rot.} & \tabheader{Trans.} & \tabheader{Rot.} & \tabheader{Trans.} & \tabheader{Rot.} \\
\toprule\bottomrule
\multirow{9}{*}[\dimexpr\dimexpr(-\aboverulesep-\belowrulesep-\cmidrulewidth)/2\relax-1.5pt\relax]{\rotatebox[origin=c]{90}{{\color{gray} Incremental / streaming}}}  & \diminish{ORB-SLAM3 (VO)} & \diminish{\cmark} & \diminish{\xmark} & \diminish{--} & \diminish{$48.66^{*}$} & \diminish{$\times$} & \diminish{$\times$} & \diminish{$\times$} & \diminish{$\times$} & \diminish{--} & \diminish{$0.59^{*}$} \\
 & \diminish{cuVSLAM (VO)} & \diminish{\cmark} & \diminish{\xmark} & \diminish{--} & \diminish{$42.39^{*}$} & \diminish{$\times$} & \diminish{$\times$} & \diminish{$\times$} & \diminish{$\times$} & -- & $0.49$ \\
& HorizonStream & \xmark & \xmark & $61.00$ & $6.09$ & $50.15$ & $11.44$ & $49.80$ & $15.49$ & $30.42$ & $6.77$ \\
 & LingBot-Map & \xmark & \xmark & -- & $27.15$ & -- & $5.35$ & -- & $9.44$ & -- & $2.18$ \\
 & STream3R & \xmark & \xmark & -- & $70.59$ & -- & $27.54$ & -- & $24.13$ & -- & $18.85$ \\
 & ABot & \xmark & \xmark & -- & $3.95$ & -- & $2.36$ & -- & $6.66$ & -- & $1.45$ \\
 & \textbf{\ourmethodvo{} (stream)} & \xmark & \xmark & $3.75$ & $1.53$ & $3.37$ & $2.34$ & $5.84$ & $2.95$ & $15.52$ & $3.02$ \\
\cmidrule{2-12}
 & \diminish{ORB-SLAM3 (VIO)} & \diminish{\cmark} & \diminish{\cmark} & \diminish{$\times$} & \diminish{$\times$} & \diminish{$\times$} & \diminish{$\times$} & \diminish{$\times$} & \diminish{$\times$} & \diminish{$\times$} & \diminish{$\times$} \\
 & BotVIO & \xmark & \cmark & $80.25$ & $36.35$ & $80.49$ & $32.52$ & $69.00$ & $91.40$ & $88.41$ & $3.50$ \\
 & \textbf{\ourmethod{} (stream)} & \xmark & \cmark & $3.42$ & $1.08$ & $2.57$ & $0.46$ & $5.19$ & $2.00$ & $14.03$ & $1.01$ \\
\midrule
\multirow{7}{*}[\dimexpr(-\aboverulesep-\belowrulesep-\cmidrulewidth)/2\relax]{\rotatebox[origin=c]{90}{{\color{gray} Batch / submap}}} & VGGT-SLAM2$^{\dagger}$ & \cmark & \xmark & $6.62$ & $1.97$ & $15.34$ & $4.81$ & $10.06$ & $6.10$ & $20.17$ & $3.26$ \\
 & MapAnything & \xmark & \xmark & $35.51$ & $13.52$ & $15.47$ & $13.26$ & $21.37$ & $11.50$ & $23.60$ & $10.46$ \\
 & VGGT-$\Omega$ & \xmark & \xmark & -- & $8.33$ & -- & $3.40$ & -- & $5.49$ & -- & $1.75$ \\
 & \textbf{\ourmethodvo{} (batch)} & \xmark & \xmark & $4.14$ & $1.27$ & $2.84$ & $1.11$ & $5.25$ & $2.11$ & $15.01$ & $2.89$ \\
\cmidrule{2-12}
 & MASt3R-Fusion & \cmark & \cmark & \diminish{$101.61^{*}$} & \diminish{$11.03^{*}$} & $1.45$ & $0.29$ & \diminish{$4.38^{*}$} & \diminish{$1.92^{*}$} & $4.18$ & $0.35$ \\
 & FoundationVIO$^{\dagger}$ & \cmark & \cmark & $7.67$ & $1.61$ & $1.37$ & $0.20$ & $4.92$ & $1.82$ & $10.41$ & $0.19$ \\
 & \textbf{\ourmethod{} (batch)} & \xmark & \cmark & $2.96$ & $1.15$ & $2.35$ & $0.47$ & $5.15$ & $1.79$ & $15.48$ & $0.47$ \\
\bottomrule
\end{NiceTabular}%
}
\par\vspace{2pt}
{\raggedright\footnotesize
$*$: failures excluded; $\times$: no completed segment; --: scale-ambiguous translation or no eligible pairs; $\dagger$: uses finetuned \ourmethod{}/\ourmethodvo frontend; $^\ddagger$: held-out dataset excluded from training.
%\lc{how can a method have $--$ for translation and $\times$ for rotation?}
% FoundationVIO uses normal settings. RPE uses eligible pairs from completed segments; no missing pair is assigned zero error.
\par}
\vspace{-12pt}

\end{table}

% Main indoor table APE

\begin{table*}[t!]
\caption{\textbf{Odometry performance on driving datasets.} APE (m) after alignments with a monocular camera and loop closures disabled. \tabheader{Opt}: iterative optimization at inference. \tabheader{IMU}: IMU enabled. \tabheader{Metric}: Odometry is in metric scale. \tabheader{C/A}: completed/attempted segments. Results are split into VO/VIO and streaming/batched, with the best- and second-best of fully completed evaluations highlighted in gray.}
\vspace{-7pt}
\label{tab:draft:driving:ape}
\centering
\resizebox{0.77\textwidth}{!}{%
\renewcommand{\arraystretch}{1.08}
\setlength{\tabcolsep}{3pt}
\begin{NiceTabular}{@{\hspace{2pt}}c@{\hspace{6pt}}l|ccc|ccc|ccc|ccc|ccc}[
            code-before =%
            % VO best
            \rectanglecolor{\bestcolor}{9-6}{9-7}
            \rectanglecolor{\bestcolor}{9-9}{9-9}
            \rectanglecolor{\bestcolor}{8-10}{8-10}
            \rectanglecolor{\bestcolor}{9-12}{9-13}
            \rectanglecolor{\bestcolor}{9-15}{9-15}
            \rectanglecolor{\bestcolor}{8-16}{8-16}
            % VO second best
            \rectanglecolor{\secondbestcolor}{5-6}{5-6}
            \rectanglecolor{\secondbestcolor}{8-7}{8-7}
            \rectanglecolor{\secondbestcolor}{5-9}{5-9}
            \rectanglecolor{\secondbestcolor}{9-10}{9-10}
            \rectanglecolor{\secondbestcolor}{5-12}{5-12}
            \rectanglecolor{\secondbestcolor}{8-13}{8-13}
            \rectanglecolor{\secondbestcolor}{5-15}{5-15}
            \rectanglecolor{\secondbestcolor}{6-16}{6-16}
            % VIO best
            \rectanglecolor{\bestcolor}{12-6}{12-7}
            \rectanglecolor{\bestcolor}{12-9}{12-10}
            \rectanglecolor{\bestcolor}{12-12}{12-13}
            \rectanglecolor{\bestcolor}{12-15}{12-16}
            % VIO second best
            \rectanglecolor{\secondbestcolor}{11-6}{11-7}
            \rectanglecolor{\secondbestcolor}{11-9}{11-10}
            \rectanglecolor{\secondbestcolor}{11-12}{11-13}
            \rectanglecolor{\secondbestcolor}{11-15}{11-16}
            % VO best
            \rectanglecolor{\bestcolor}{16-6}{16-7}
            \rectanglecolor{\bestcolor}{16-9}{16-10}
            \rectanglecolor{\bestcolor}{16-12}{16-13}
            \rectanglecolor{\bestcolor}{16-15}{16-16}
            % VO second best
            \rectanglecolor{\secondbestcolor}{13-6}{13-7}
            \rectanglecolor{\secondbestcolor}{13-9}{13-10}
            \rectanglecolor{\secondbestcolor}{13-12}{13-13}
            \rectanglecolor{\secondbestcolor}{13-15}{13-15}
            \rectanglecolor{\secondbestcolor}{15-16}{15-16}
            % VIO second best
            \rectanglecolor{\bestcolor}{19-6}{19-7}
            \rectanglecolor{\bestcolor}{18-9}{18-10}
            \rectanglecolor{\bestcolor}{19-12}{19-13}
            \rectanglecolor{\bestcolor}{18-15}{18-15}
            \rectanglecolor{\bestcolor}{19-16}{19-16}
            % VIO second best
            \rectanglecolor{\secondbestcolor}{17-6}{17-7}
            \rectanglecolor{\secondbestcolor}{19-9}{19-10}
            \rectanglecolor{\secondbestcolor}{17-12}{17-13}
            \rectanglecolor{\secondbestcolor}{19-15}{19-15}
            \rectanglecolor{\secondbestcolor}{18-16}{18-16}
        ]
\toprule
\multicolumn{2}{c|}{\tabheader{Method}} & \tabheader{IMU} & \tabheader{Opt.} & \tabheader{Metric} & \multicolumn{3}{c|}{\tabheader{\ourdataset}} & \multicolumn{3}{c|}{\tabheader{KITTI-360}} & \multicolumn{3}{c|}{\tabheader{TartanDrive 2}} & \multicolumn{3}{c}{\tabheader{KITTI Odom.$^\ddagger$}} \\
 &  &  &  &  & \tabheader{SE(3)} & \tabheader{Sim(3)} & \tabheader{C/A} & \tabheader{SE(3)} & \tabheader{Sim(3)} & \tabheader{C/A} & \tabheader{SE(3)} & \tabheader{Sim(3)} & \tabheader{C/A} & \tabheader{SE(3)} & \tabheader{Sim(3)} & \tabheader{C/A} \\
\toprule\bottomrule
\multirow{10}{*}{\rotatebox[origin=c]{90}{{\color{gray}\scriptsize Incremental / streaming}}} & \diminish{ORB-SLAM3 (VO)} & \diminish{\xmark} & \diminish{\cmark} & \diminish{\xmark} & \diminish{--} & \diminish{$97.56^{*}$} & \diminish{$6/17$} & \diminish{--} & \diminish{$\times$} & \diminish{$0/2$} & \diminish{--} & \diminish{$0.03^{*}$} & \diminish{$2/4$} & \diminish{--} & \diminish{$28.19^{*}$} & \diminish{$15/17$} \\
 & \diminish{cuVSLAM (VO)} & \diminish{\xmark} & \diminish{\cmark} & \diminish{\xmark} & \diminish{--} & \diminish{$19.01^{*}$} & \diminish{$6/17$} & \diminish{--} & \diminish{$\times$} & \diminish{$0/2$} & \diminish{--} & \diminish{$0.03^{*}$} & \diminish{$2/4$} & \diminish{--} & \diminish{$31.20$} & \diminish{$17/17$} \\
 & HorizonStream & \xmark & \xmark & \cmark & $107.24$ & $37.21$ & $17/17$ & $326.33$ & $160.36$ & $2/2$ & $123.06$ & $45.78$ & $4/4$ & $92.65$ & $31.56$ & $17/17$ \\
 & LingBot-Map & \xmark & \xmark & \xmark & -- & $84.87$ & $17/17$ & -- & $32.37$ & $2/2$ & -- & $21.94$ & $4/4$ & -- & $16.05$ & $17/17$ \\
 & STream3R & \xmark & \xmark & \xmark & -- & $114.04$ & $17/17$ & -- & $226.45$ & $2/2$ & -- & $103.60$ & $4/4$ & -- & $185.18$ & $17/17$ \\
 & ABot & \xmark & \xmark & \xmark & -- & $25.54$ & $17/17$ & -- & $19.84$ & $2/2$ & -- & $14.00$ & $4/4$ & -- & $7.97$ & $17/17$ \\
 & \textbf{\ourmethodvo{} (stream)} & \xmark & \xmark & \cmark & $8.97$ & $7.91$ & $17/17$ & $23.55$ & $22.83$ & $2/2$ & $7.97$ & $7.60$ & $4/4$ & $40.43$ & $18.10$ & $17/17$ \\\cmidrule{2-17}
 & ORB-SLAM3 (VIO) & \cmark & \cmark & \cmark
 & $\times$ & $\times$ & $0/17$
 & $\times$ & $\times$ & $0/2$
 & $\times$ & $\times$ & $0/4$
 & $\times$ & $\times$ & $0/17$ \\
 & BotVIO & \cmark & \xmark & \cmark & $111.89$ & $69.39$ & $17/17$ & $221.11$ & $175.93$ & $2/2$ & $103.42$ & $90.64$ & $4/4$ & $183.10$ & $29.27$ & $17/17$ \\
 & \textbf{\ourmethod{} (stream)} & \cmark & \xmark & \cmark & $5.90$ & $5.22$ & $17/17$ & $6.96$ & $6.48$ & $2/2$ & $3.73$ & $3.39$ & $4/4$ & $30.88$ & $9.28$ & $17/17$ \\
\midrule
\multirow{6}{*}{\rotatebox[origin=c]{90}{{\color{gray}\scriptsize Batch / submap}}} & VGGT-SLAM2$^{\dagger}$ & \xmark & \cmark & \cmark & $9.99$ & $6.15$ & $17/17$ & $34.33$ & $16.04$ & $2/2$ & $11.16$ & $8.77$ & $4/4$ & $45.10$ & $17.13$ & $17/17$ \\
 & MapAnything & \xmark & \xmark & \cmark & $94.26$ & $36.05$ & $17/17$ & $200.40$ & $156.58$ & $2/2$ & $34.02$ & $15.44$ & $4/4$ & $76.65$ & $55.20$ & $17/17$ \\
 & VGGT-Omega & \xmark & \xmark & \xmark & -- & $34.91$ & $17/17$ & -- & $52.41$ & $2/2$ & -- & $20.88$ & $4/4$ & -- & $16.76$ & $17/17$ \\
 & \textbf{\ourmethodvo{} (batch)} & \xmark & \xmark & \cmark & $7.38$ & $5.09$ & $17/17$ & $12.89$ & $12.00$ & $2/2$ & $4.90$ & $3.97$ & $4/4$ & $37.66$ & $16.35$ & $17/17$ \\\cmidrule{2-17}
 & FoundationVIO$^{\dagger}$ & \cmark & \cmark & \cmark & $7.97$ & $6.03$ & $17/17$ & $9.92$ & $8.07$ & $2/2$ & $4.12$ & $3.60$ & $4/4$ & $52.80$ & $25.02$ & $17/17$ \\
  & MASt3R-Fusion & \cmark & \cmark & \cmark & $135.71^{*}$ & $74.18^{*}$ & $6/17$ & $2.96$ & $2.33$ & $2/2$ & $1.53^{*}$ & $1.34^{*}$ & $2/4$ & $8.79$ & $4.88$ & $17/17$ \\
 & \textbf{\ourmethod{} (batch)} & \cmark & \xmark & \cmark & $4.94$ & $4.65$ & $17/17$ & $6.34$ & $5.89$ & $2/2$ & $3.29$ & $2.60$ & $4/4$ & $31.98$ & $4.26$ & $17/17$ \\
\bottomrule
\end{NiceTabular}%
}
\par\vspace{2pt}
{\raggedright
\footnotesize
 $*$ excludes failures; $\times$ no successful segment;
 -- is inapplicable or no eligible pairs;
% $?$: unresolved evidence;
$\dagger$ uses finetuned \ourmethod{}/\ourmethodvo frontend; $^\ddagger$ is a held-out dataset. 
\par}
\end{table*}

% Main indoor table RPE
\begin{table}[t!]
\caption{\textbf{Performance on indoor and walking datasets.} RPE per 1 s with a monocular camera and loop closures disabled. \tabheader{Opt}: iterative optimization at inference. \tabheader{IMU}: IMU enabled. Mean relative errors are reported: translation in m and rotation in $^\circ$. Completion counts are in the corresponding APE table. Best and second-best results among fully completed evaluations are highlighted in gray, separately within each VO/VIO and streaming/batched block.}
\vspace{-7pt}
\label{tab:main:rpe:indoor_walking}
\centering
\resizebox{\columnwidth}{!}{%
\renewcommand{\arraystretch}{1.08}
\setlength{\tabcolsep}{3pt}
\begin{NiceTabular}{@{\hspace{2pt}}c@{\hspace{6pt}}l|cc|cc|cc|cc|cc|cc}[code-before={%
\rectanglecolor{\secondbestcolor}{5-5}{5-5}
\rectanglecolor{\bestcolor}{5-7}{5-7}
\rectanglecolor{\secondbestcolor}{5-9}{5-9}
\rectanglecolor{\secondbestcolor}{5-11}{5-11}
\rectanglecolor{\secondbestcolor}{5-14}{5-14}
\rectanglecolor{\secondbestcolor}{6-10}{6-10}
\rectanglecolor{\secondbestcolor}{8-6}{8-6}
\rectanglecolor{\secondbestcolor}{8-8}{8-8}
\rectanglecolor{\secondbestcolor}{9-13}{9-13}
\rectanglecolor{\bestcolor}{8-12}{8-12}
\rectanglecolor{\bestcolor}{9-5}{9-5}
\rectanglecolor{\bestcolor}{9-6}{9-6}
\rectanglecolor{\bestcolor}{9-7}{9-7}
\rectanglecolor{\bestcolor}{9-8}{9-8}
\rectanglecolor{\bestcolor}{9-9}{9-9}
\rectanglecolor{\bestcolor}{9-10}{9-10}
\rectanglecolor{\bestcolor}{9-11}{9-11}
\rectanglecolor{\bestcolor}{5-13}{5-13}
\rectanglecolor{\secondbestcolor}{9-12}{9-12}
\rectanglecolor{\bestcolor}{6-14}{6-14}
\rectanglecolor{\bestcolor}{8-14}{8-14}
%%%
\rectanglecolor{\secondbestcolor}{10-11}{10-11}
\rectanglecolor{\secondbestcolor}{10-12}{10-12}
\rectanglecolor{\secondbestcolor}{11-5}{11-5}
\rectanglecolor{\secondbestcolor}{11-6}{11-6}
\rectanglecolor{\secondbestcolor}{11-7}{11-7}
\rectanglecolor{\secondbestcolor}{11-8}{11-8}
\rectanglecolor{\secondbestcolor}{11-13}{11-13}
\rectanglecolor{\secondbestcolor}{11-14}{11-14}
\rectanglecolor{\bestcolor}{12-5}{12-5}
\rectanglecolor{\bestcolor}{12-6}{12-6}
\rectanglecolor{\bestcolor}{12-7}{12-7}
\rectanglecolor{\bestcolor}{12-8}{12-8}
\rectanglecolor{\bestcolor}{12-9}{12-9}
\rectanglecolor{\bestcolor}{12-10}{12-10}
\rectanglecolor{\bestcolor}{12-11}{12-11}
\rectanglecolor{\bestcolor}{12-12}{12-12}
\rectanglecolor{\bestcolor}{12-13}{12-13}
\rectanglecolor{\bestcolor}{12-14}{12-14}
%%%
\rectanglecolor{\bestcolor}{13-5}{13-5}
\rectanglecolor{\secondbestcolor}{13-7}{13-7}
\rectanglecolor{\bestcolor}{13-9}{12-9}
\rectanglecolor{\secondbestcolor}{13-10}{13-10}
\rectanglecolor{\secondbestcolor}{13-11}{13-11}
\rectanglecolor{\secondbestcolor}{13-13}{13-13}
\rectanglecolor{\secondbestcolor}{13-14}{13-14}
\rectanglecolor{\secondbestcolor}{14-9}{14-9}
\rectanglecolor{\bestcolor}{14-13}{14-13}
\rectanglecolor{\secondbestcolor}{15-6}{15-6}
\rectanglecolor{\secondbestcolor}{15-8}{15-8}
\rectanglecolor{\secondbestcolor}{15-12}{15-12}
\rectanglecolor{\bestcolor}{15-14}{15-14}
\rectanglecolor{\secondbestcolor}{16-5}{16-5}
\rectanglecolor{\bestcolor}{16-6}{16-6}
\rectanglecolor{\bestcolor}{16-7}{16-7}
\rectanglecolor{\bestcolor}{16-8}{16-8}
\rectanglecolor{\bestcolor}{16-9}{16-9}
\rectanglecolor{\bestcolor}{16-10}{16-10}
\rectanglecolor{\bestcolor}{16-11}{16-11}
\rectanglecolor{\bestcolor}{16-12}{16-12}
\rectanglecolor{\secondbestcolor}{16-13}{16-13}
%%%
\rectanglecolor{\secondbestcolor}{17-11}{17-11}
\rectanglecolor{\secondbestcolor}{18-5}{18-5}
\rectanglecolor{\secondbestcolor}{18-6}{18-6}
\rectanglecolor{\bestcolor}{18-7}{18-7}
\rectanglecolor{\bestcolor}{18-8}{18-8}
\rectanglecolor{\secondbestcolor}{18-9}{18-9}
\rectanglecolor{\bestcolor}{18-10}{18-10}
\rectanglecolor{\bestcolor}{18-11}{18-11}
\rectanglecolor{\bestcolor}{18-12}{18-12}
\rectanglecolor{\bestcolor}{18-13}{18-13}
\rectanglecolor{\secondbestcolor}{18-14}{18-14}
\rectanglecolor{\bestcolor}{19-5}{19-5}
\rectanglecolor{\bestcolor}{19-6}{19-6}
\rectanglecolor{\secondbestcolor}{19-7}{19-7}
\rectanglecolor{\secondbestcolor}{19-8}{19-8}
\rectanglecolor{\bestcolor}{19-9}{19-9}
\rectanglecolor{\secondbestcolor}{19-10}{19-10}
\rectanglecolor{\secondbestcolor}{19-12}{19-12}
\rectanglecolor{\secondbestcolor}{19-13}{19-13}
\rectanglecolor{\bestcolor}{19-14}{19-14}
}]
\toprule
\multicolumn{2}{c|}{\tabheader{Method}} & \tabheader{Opt.} & \tabheader{IMU} & \multicolumn{2}{c|}{\tabheader{GrandTour}} & \multicolumn{2}{c|}{\tabheader{Oxford Spires}} & \multicolumn{2}{c|}{\tabheader{ADT}} & \multicolumn{2}{c|}{\tabheader{ScaRF$^\ddagger$}} & \multicolumn{2}{c}{\tabheader{OpenLORIS$^\ddagger$}} \\
 &  &  &  & \tabheader{Trans.} & \tabheader{Rot.} & \tabheader{Trans.} & \tabheader{Rot.} & \tabheader{Trans.} & \tabheader{Rot.} & \tabheader{Trans.} & \tabheader{Rot.} & \tabheader{Trans.} & \tabheader{Rot.} \\
\toprule\bottomrule
\multirow{9}{*}[\dimexpr\dimexpr(-\aboverulesep-\belowrulesep-\cmidrulewidth)/2\relax-1.5pt\relax]{\rotatebox[origin=c]{90}{{\color{gray} Incremental / streaming}}} & \diminish{ORB-SLAM3 (VO)} & \diminish{\cmark} & \diminish{\xmark} & \diminish{--} & \diminish{$0.26^{*}$} & \diminish{$\times$} & \diminish{$\times$} & \diminish{--} & \diminish{$0.32^{*}$} & \diminish{--} & \diminish{$0.26^{*}$} & \diminish{--} & \diminish{$0.74^{*}$} \\
 & \diminish{cuVSLAM (VO)} & \diminish{\cmark} & \diminish{\xmark} & \diminish{--} & \diminish{$0.23^{*}$} & \diminish{$\times$} & \diminish{$\times$} & \diminish{--} & \diminish{$0.47^{*}$} & \diminish{--} & \diminish{$0.36^{*}$} & \diminish{--} & \diminish{$0.86^{*}$} \\
 & HorizonStream & \xmark & \xmark & $0.41$ & $0.64$ & $0.11$ & $0.50$ & $0.03$ & $0.47$ & $0.10$ & $0.93$ & $0.05$ & $0.82$ \\
 & LingBot-Map & \xmark & \xmark & -- & $1.31$ & -- & $4.19$ & -- & $0.33$ & -- & $2.87$ & -- & $0.67$ \\
 & STream3R & \xmark & \xmark & -- & $14.89$ & -- & $16.76$ & -- & $1.38$ & -- & $33.36$ & -- & $12.09$ \\
 & ABot & \xmark & \xmark & -- & $0.38$ & -- & $0.42$ & -- & $0.41$ & -- & $0.53$ & -- & $0.67$ \\
 & \textbf{\ourmethodvo{} (stream)} & \xmark & \xmark & $0.08$ & $0.32$ & $0.11$ & $0.27$ & $0.01$ & $0.25$ & $0.09$ & $0.71$ & $0.09$ & $1.36$ \\
\cmidrule{2-14}
 & ORB-SLAM3 (VIO) & \cmark & \cmark
 & \diminish{$0.56^{*}$} & \diminish{$0.26^{*}$} % GrandTour
 & \diminish{$0.06^{*}$} & \diminish{$0.12^{*}$} % Oxford Spires
 & \diminish{$0.22^{*}$} & \diminish{$1.71^{*}$} % ADT
 & $0.03$ & $0.20$                             % ScaRF
 & \diminish{$\times$} & \diminish{$\times$} \\ % OpenLORIS
 & BotVIO & \xmark & \cmark & $0.43$ & $2.90$ & $1.07$ & $7.47$ & \diminish{$0.51^{*}$} & \diminish{$13.45^{*}$} & $0.50$ & $13.45$ & $0.13$ & $4.11$ \\
 & \textbf{\ourmethod{} (stream)} & \xmark & \cmark & $0.08$ & $0.26$ & $0.11$ & $0.11$ & $0.01$ & $0.19$ & $0.09$ & $0.24$ & $0.09$ & $0.45$ \\
\midrule
\multirow{7}{*}[\dimexpr(-\aboverulesep-\belowrulesep-\cmidrulewidth)/2\relax]{\rotatebox[origin=c]{90}{{\color{gray} Batch / submap}}} & VGGT-SLAM2$^{\dagger}$ & \cmark & \xmark & $0.09$ & $0.58$ & $0.17$ & $0.78$ & $0.01$ & $0.25$ & $0.09$ & $1.17$ & $0.12$ & $0.77$ \\
 & MapAnything  & \xmark & \xmark & $0.23$ & $0.58$ & $0.18$ & $0.46$ & $0.03$ & $0.47$ & $0.17$ & $1.14$ & $0.06$ & $0.91$ \\
 & VGGT-$\Omega$  & \xmark & \xmark & -- & $0.35$ & -- & $0.25$ & -- & $0.27$ & -- & $0.56$ & -- & $0.51$ \\
 & \textbf{\ourmethodvo{} (batch)} & \xmark & \xmark & $0.10$ & $0.26$ & $0.09$ & $0.18$ & $0.01$ & $0.21$ & $0.08$ & $0.40$ & $0.12$ & $0.79$ \\
\cmidrule{2-14}
 & MASt3R-Fusion & \cmark & \cmark & $0.17$ & $0.29$ & $0.16$ & $0.19$ & \diminish{$0.10^{*}$} & \diminish{$0.21^{*}$} & $0.06$ & $0.26$ & \diminish{$0.52^{*}$} & \diminish{$0.73^{*}$} \\
 & FoundationVIO$^{\dagger}$ & \cmark & \cmark & $0.13$ & $0.26$ & $0.07$ & $0.11$ & $0.02$ & $0.18$ & $0.03$ & $0.19$ & $0.07$ & $0.46$ \\
 & \textbf{\ourmethod{} (batch)} & \xmark & \cmark & $0.08$ & $0.25$ & $0.10$ & $0.12$ & $0.01$ & $0.19$ & $0.08$ & $0.22$ & $0.11$ & $0.43$ \\
\bottomrule
\end{NiceTabular}%
}
\par\vspace{2pt}
{\raggedright\footnotesize
$*$: failures excluded; $\times$: no completed segment; --: scale-ambiguous translation or no eligible pairs; $\dagger$: uses finetuned \ourmethod{}/\ourmethodvo frontend; $^\ddagger$: held-out dataset excluded from training.% FoundationVIO uses normal settings. RPE uses eligible pairs from completed segments; no missing pair is assigned zero error.
\par}
\vspace{-12pt}
\end{table}

% Main indoor table APE

\begin{table*}[t!]
\caption{\textbf{Odometry performance on indoor and walking datasets.} APE (m) after alignments with a monocular camera and loop closures disabled. \tabheader{Opt}: iterative optimization at inference. \tabheader{IMU}: IMU enabled. \tabheader{Metric}: Odometry is metric. \tabheader{C/A}: completed/attempted segments. Results are split into VO/VIO and streaming/batched, with the best- and second-best of fully completed evaluations highlighted in gray.}
\vspace{-7pt}
\label{tab:draft:indoor_walking:ape}
\centering
\resizebox{\textwidth}{!}{%
\begin{NiceTabular}{ll|ccc|ccc|ccc|ccc|ccc|ccc}[
            code-before =%
            % VO best
            \rectanglecolor{\bestcolor}{9-6}{9-7}
            \rectanglecolor{\bestcolor}{9-9}{9-10}
            \rectanglecolor{\bestcolor}{9-12}{9-12}
            \rectanglecolor{\bestcolor}{6-13}{6-13}
            \rectanglecolor{\bestcolor}{9-15}{9-15}
            \rectanglecolor{\bestcolor}{8-16}{8-16}
            \rectanglecolor{\bestcolor}{5-18}{5-18}
            \rectanglecolor{\bestcolor}{6-19}{6-19}
            % VO second best
            \rectanglecolor{\secondbestcolor}{5-6}{5-6}
            \rectanglecolor{\secondbestcolor}{6-7}{6-7}
            \rectanglecolor{\secondbestcolor}{5-9}{5-9}
            \rectanglecolor{\secondbestcolor}{8-10}{8-10}
            \rectanglecolor{\secondbestcolor}{5-12}{5-12}
            \rectanglecolor{\secondbestcolor}{9-13}{9-13}
            \rectanglecolor{\secondbestcolor}{5-15}{5-15}
            \rectanglecolor{\secondbestcolor}{9-16}{9-16}
            \rectanglecolor{\secondbestcolor}{9-18}{9-18}
            \rectanglecolor{\secondbestcolor}{5-19}{5-19}
            \rectanglecolor{\secondbestcolor}{8-19}{8-19}
            % VIO best
            \rectanglecolor{\bestcolor}{12-6}{12-7}
            \rectanglecolor{\bestcolor}{12-9}{12-10}
            \rectanglecolor{\bestcolor}{12-12}{12-13}
            \rectanglecolor{\bestcolor}{10-15}{10-16}
            \rectanglecolor{\bestcolor}{12-18}{12-19}
            % VIO second best
            \rectanglecolor{\secondbestcolor}{11-6}{11-7}
            \rectanglecolor{\secondbestcolor}{11-9}{11-10}
            \rectanglecolor{\secondbestcolor}{12-15}{12-16}
            \rectanglecolor{\secondbestcolor}{11-18}{11-19}
            % VO best
            \rectanglecolor{\bestcolor}{13-6}{13-6}
            \rectanglecolor{\bestcolor}{16-7}{16-7}
            \rectanglecolor{\bestcolor}{16-9}{16-10}
            \rectanglecolor{\bestcolor}{13-12}{13-13}
            \rectanglecolor{\bestcolor}{16-13}{16-13}
            \rectanglecolor{\bestcolor}{16-15}{16-16}
            \rectanglecolor{\bestcolor}{14-18}{14-19}
            % VO second best
            \rectanglecolor{\secondbestcolor}{16-6}{16-6}
            \rectanglecolor{\secondbestcolor}{13-7}{13-7}
            \rectanglecolor{\secondbestcolor}{13-9}{13-10}
            \rectanglecolor{\secondbestcolor}{16-12}{16-12}
            \rectanglecolor{\secondbestcolor}{13-15}{13-16}
            \rectanglecolor{\secondbestcolor}{13-18}{13-19}
            \rectanglecolor{\secondbestcolor}{16-18}{16-19}
            %% VIO second best
            \rectanglecolor{\bestcolor}{18-6}{18-7}
            \rectanglecolor{\bestcolor}{18-9}{18-10}
            \rectanglecolor{\bestcolor}{19-12}{19-13}
            \rectanglecolor{\bestcolor}{18-15}{18-16}
            \rectanglecolor{\bestcolor}{17-18}{17-19}
            %% VIO second best
            \rectanglecolor{\secondbestcolor}{19-6}{19-7}
            \rectanglecolor{\secondbestcolor}{17-9}{17-10}
            \rectanglecolor{\secondbestcolor}{19-10}{19-10}
            \rectanglecolor{\secondbestcolor}{17-12}{17-13}
            \rectanglecolor{\secondbestcolor}{17-15}{17-16}
            \rectanglecolor{\secondbestcolor}{19-18}{19-19}
        ]
\toprule
\multicolumn{2}{c|}{\tabheader{Method}} & \tabheader{IMU} & \tabheader{Opt.} & \tabheader{Metric} & \multicolumn{3}{c|}{\tabheader{GrandTour}} & \multicolumn{3}{c|}{\tabheader{Oxford Spires}} & \multicolumn{3}{c|}{\tabheader{ADT}} & \multicolumn{3}{c|}{\tabheader{ScaRF$^\ddagger$}} & \multicolumn{3}{c}{\tabheader{OpenLORIS$^\ddagger$}} \\
 &  &  &  &  & \tabheader{SE(3)} & \tabheader{Sim(3)} & \tabheader{C/A} & \tabheader{SE(3)} & \tabheader{Sim(3)} & \tabheader{C/A} & \tabheader{SE(3)} & \tabheader{Sim(3)} & \tabheader{C/A} & \tabheader{SE(3)} & \tabheader{Sim(3)} & \tabheader{C/A} & \tabheader{SE(3)} & \tabheader{Sim(3)} & \tabheader{C/A} \\
\toprule\bottomrule
\multirow{11}{*}{\rotatebox[origin=c]{90}{{\color{gray}Incremental / streaming}}} & ORB-SLAM3 (VO) & \xmark & \cmark & \xmark & -- & $1.47^{*}$ & $2/8$ & -- & $\times$ & $0/3$ & -- & $0.04^{*}$ & $11/16$ & -- & $1.36^{*}$ & $2/3$ & -- & $0.03^{*}$ & $4/5$ \\
 & cuVSLAM (VO) & \xmark & \cmark & \xmark & -- & $3.57^{*}$ & $4/8$ & -- & $\times$ & $0/3$ & -- & $0.12^{*}$ & $10/16$ & -- & $1.62^{*}$ & $1/3$ & -- & $0.04^{*}$ & $4/5$ \\
 & HorizonStream & \xmark & \xmark & \cmark & $16.67$ & $5.47$ & $8/8$ & $13.83$ & $13.08$ & $3/3$ & $0.23$ & $0.16$ & $16/16$ & $0.99$ & $0.99$ & $3/3$ & $0.16$ & $0.06$ & $5/5$ \\
 & LingBot-Map & \xmark & \xmark & \xmark & -- & $2.45$ & $8/8$ & -- & $10.83$ & $3/3$ & -- & $0.03$ & $16/16$ & -- & $1.46$ & $3/3$ & -- & $0.04$ & $5/5$ \\
 & STream3R & \xmark & \xmark & \xmark & -- & $16.16$ & $8/8$ & -- & $34.51$ & $3/3$ & -- & $0.12$ & $16/16$ & -- & $5.34$ & $3/3$ & -- & $0.64$ & $5/5$ \\
 & ABot & \xmark & \xmark & \xmark & -- & $2.49$ & $8/8$ & -- & $5.69$ & $3/3$ & -- & $0.17$ & $16/16$ & -- & $0.52$ & $3/3$ & -- & $0.06$ & $5/5$ \\
 & \textbf{\ourmethodvo{} (stream)} & \xmark & \xmark & \cmark & $2.38$ & $1.73$ & $8/8$ & $2.67$ & $2.04$ & $3/3$ & $0.06$ & $0.06$ & $16/16$ & $0.95$ & $0.83$ & $3/3$ & $0.23$ & $0.15$ & $5/5$ \\
\cmidrule{2-20}
 & ORB-SLAM3 (VIO) & \cmark & \cmark & \cmark
 & $28.49^{*}$ & $10.27^{*}$ & $2/8$ % GrandTour
 & $3.26^{*}$ & $1.99^{*}$ & $1/3$   % Oxford Spires
 & $0.57^{*}$ & $0.26^{*}$ & $1/16$  % ADT
 & $0.30$ & $0.23$ & $3/3$           % ScaRF
 & $\times$ & $\times$ & $0/5$ \\    % OpenLORIS
 & BotVIO & \cmark & \xmark & \cmark & $15.79$ & $14.79$ & $8/8$ & $28.19$ & $25.55$ & $3/3$ & $5.46^{*}$ & $0.21^{*}$ & $1/16$ & $6.32$ & $4.79$ & $3/3$ & $0.55$ & $0.33$ & $5/5$ \\
 & \textbf{\ourmethod{} (stream)} & \cmark & \xmark & \cmark & $2.13$ & $1.55$ & $8/8$ & $2.45$ & $1.82$ & $3/3$ & $0.05$ & $0.05$ & $16/16$ & $0.85$ & $0.69$ & $3/3$ & $0.22$ & $0.15$ & $5/5$ \\
\midrule
\multirow{6}{*}{\rotatebox[origin=c]{90}{{\color{gray}Batch / submap}}} & VGGT-SLAM2$^\dagger$ & \xmark & \cmark & \cmark & $2.96$ & $1.81$ & $8/8$ & $4.31$ & $3.42$ & $3/3$ & $0.04$ & $0.04$ & $16/16$ & $0.74$ & $0.66$ & $3/3$ & $0.40$ & $0.11$ & $5/5$ \\
 & MapAnything & \xmark & \xmark & \cmark & $8.89$ & $3.79$ & $8/8$ & $6.39$ & $5.85$ & $3/3$ & $0.13$ & $0.10$ & $16/16$ & $1.56$ & $1.01$ & $3/3$ & $0.14$ & $0.08$ & $5/5$ \\
 & VGGT-Omega & \xmark & \xmark & \xmark & -- & $5.54$ & $8/8$ & -- & $14.08$ & $3/3$ & -- & $0.06$ & $16/16$ & -- & $1.35$ & $3/3$ & -- & $0.12$ & $5/5$ \\
 & \textbf{\ourmethodvo{} (batch)} & \xmark & \xmark & \cmark & $3.06$ & $1.68$ & $8/8$ & $2.85$ & $2.26$ & $3/3$ & $0.05$ & $0.04$ & $16/16$ & $0.71$ & $0.57$ & $3/3$ & $0.40$ & $0.11$ & $5/5$ \\
\cmidrule{2-20}
 & FoundationVIO$^\dagger$ & \cmark & \cmark & \cmark & $5.84$ & $3.41$ & $8/8$ & $2.17$ & $1.90$ & $3/3$ & $0.07$ & $0.06$ & $16/16$ & $0.35$ & $0.35$ & $3/3$ & $0.17$ & $0.11$ & $5/5$ \\
 & MASt3R-Fusion & \cmark & \cmark & \cmark & $1.72$ & $1.60$ & $8/8$ & $1.76$ & $1.75$ & $3/3$ & $0.17^{*}$ & $0.13^{*}$ & $10/16$ & $0.25$ & $0.23$ & $3/3$ & $1.28^{*}$ & $1.25^{*}$ & $1/5$ \\
 & \textbf{\ourmethod{} (batch)} & \cmark & \xmark & \cmark & $2.30$ & $1.67$ & $8/8$ & $2.44$ & $1.90$ & $3/3$ & $0.05$ & $0.04$ & $16/16$ & $0.57$ & $0.49$ & $3/3$ & $0.38$ & $0.12$ & $5/5$ \\
\bottomrule
\end{NiceTabular}%
}
\par\vspace{2pt}
{\raggedright
\footnotesize
 $*$ excludes failures; $\times$ no successful segment;
 -- is inapplicable or no eligible pairs;
% $?$: unresolved evidence;
$\dagger$ uses finetuned \ourmethod{}/\ourmethodvo frontend; $^\ddagger$ is held-out dataset. 
\par}
\vspace{-12pt}
\end{table*}

\subsection{Baselines}

Baselines are grouped along the two axes used in our tables: streaming or batch processing, and visual-only (VO) or visual-inertial (VIO) input. All methods run with a monocular camera and with loop closures (LC) disabled.

\noindent\textbf{Streaming.} We compare \ourmethodvo to the classical geometric methods ORB-SLAM3~\cite{campos2021orb} and cuVSLAM~\cite{korovko2025cuvslam}, and also evaluate it against the learned methods HorizonStream~\cite{cheng2026horizonstream}, LingBot-Map~\cite{chen2026lingbotmap}, STream3R~\cite{lan2025stream3r}, and ABot~\cite{han2026revisiting}. In the VIO setting, we evaluate \ourmethod against BotVIO~\cite{wei2025botvio}, a lightweight self-supervised VIO that fuses visual and inertial features with cross-attention.

\noindent\textbf{Batch.} When using bidirectional multi-view attention on batches of images, we compare \ourmethodvo with VGGT-SLAM2~\cite{maggio2026vggt}, MapAnything~\cite{keetha2026mapanythinguniversalfeedforwardmetric}, and \VGGTO~\cite{wang2026vggtomega}, with odometry stitched as described in~\cref{sec:method}. With the IMU, we compare \ourmethod to MASt3R-Fusion~\cite{zhou2025mast3r} and FoundationVIO~\cite{rhee2026foundationvio}, which solve factor-graph optimization problems with Sim(3) and IMU pre-integration factors. For VGGT-SLAM2 and FoundationVIO, we use the \ourmethodvo frontend for fair comparisons. %\lc{is this for fairness? unclear why this is the best option}. 
In FoundationVIO, we changed the linear solver in iSAM2~\cite{kaess2012isam2} from the default Cholesky factorization to QR decomposition, which is necessary for the method to complete the longer driving sequences. %The method is otherwise unchanged.

\iffalse
\noindent\textbf{Experiments.} We perform three evaluations: starting with driving datasets, then indoor and walking datasets, and finally online, onboard a vehicle for on- and off-road driving.
\fi 

\noindent\textbf{Metrics.}
We compute the mean relative pose error (RPE) over 100\,m intervals for driving datasets, and 1\,s intervals for indoor and walking datasets, and report an absolute position error (APE) under SE(3) and Sim(3) alignments % to assess metric scale estimation 
using \texttt{EVO}~\cite{grupp2017evo}. 
%A given recording is evaluated in segments if the dataset contains IMU dropouts; otherwise, the complete path is evaluated. 
We average each metric over the completed segments of a recording, and then over the recordings of a dataset. A segment is completed if the method initializes within the first half of the segment, leaves no gap longer than 2\,s without a pose estimate, reaches an APE no greater than the length of the reference path, and returns a pose for at least 95\% of its frames.
%Completed/attempted segments are reported as C/A, with $^*$ marking averages over a subset.
%Since SE(3) alignment does not correct scale, the gap between SE(3) and Sim(3) APE reflects the metric scale error. 
%Finally, we compute APE and RPE with the evo odometry and SLAM library~\cite{grupp2017evo}.

\begin{figure}[t!]\label{sec:eval:qual2d}
    \centering
    \includegraphics[width=0.8\linewidth]{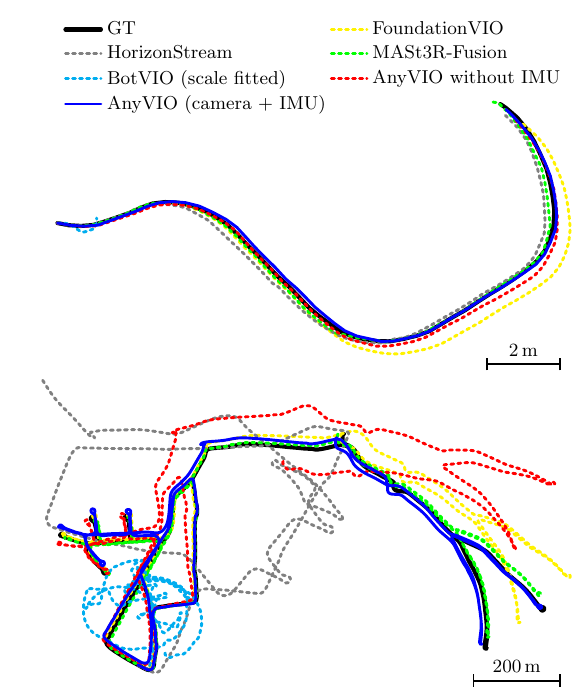}
    \vspace{-10pt}
    
    \caption{\textbf{Qualitative comparison.} Single odometry sequences from ScaRF \cite{zhang2026scarf} (top) and KITTI-360~\cite{Liao2022PAMI} (bottom). \ourmethod has the lowest error at the end of both sequences. }
    \label{fig:evalgrid}
    \vspace{-12pt}
\end{figure}

%\lc{overall, putting both RPE and APE tables might not be feasible with the page budget we have. should we push some of those to appendix?}
\subsection{Driving Evaluations}\label{sec:eval:driving}
\cref{tab:draft:driving:ape,tab:main:rpe} report results on the driving datasets. Among streaming methods, \ourmethod has the lowest SE(3) APE and translational RPE on every dataset and completes every segment, whereas ORB-SLAM3~\cite{ORBSLAM3_source} and cuVSLAM~\cite{korovko2025cuvslam} fail on most \ourdataset segments and on all of KITTI-360 \cite{Liao2022PAMI}, as seen in \cref{sec:eval:qual2d}. Without the IMU, \ourmethodvo is the most accurate metric streaming VO. After Sim(3) alignment, it is more accurate than the non-metric ABot \cite{han2026revisiting} on \ourdataset and TartanDrive~2 \cite{sivaprakasam2024tartandrive} and less accurate on the two KITTI datasets; ABot's training data includes the Virtual KITTI \cite{cabon2020vkitti2} scenes, which ours does not. The IMU lowers the APE and RPE on every dataset, mainly by reducing rotational drift. With KITTI Odometry \cite{geiger2013vision} \cite{Geiger2012CVPR}, which is not used for training, a large gap remains between SE(3) and Sim(3) APE, so the metric scale transfers only partly to unseen data.

In the batch setting, \ourmethod has a lower APE on every dataset than VGGT-SLAM2 \cite{maggio2026vggt} and FoundationVIO \cite{rhee2026foundationvio}, which use the same frontend. The optimization-based VIO methods keep a lower RPE on the two KITTI datasets, and MASt3R-Fusion \cite{zhou2025mast3r} also a lower SE(3) APE there, but it completes only 6 of 17 \ourdataset segments and its cost grows over time (\cref{tab:runtime:scaling}), whereas \ourmethod uses only forward passes. For \ourmethod, batch processing with bidirectional attention is generally more accurate than the equivalent streaming model, but the difference is marginal both for VO and VIO. 

\subsection{Indoor and Walking Evaluations}\label{sec:eval:indoor}
\cref{tab:draft:indoor_walking:ape,tab:main:rpe:indoor_walking} report results on the indoor and walking datasets. In non-driving environments, we observe that classical methods are quite accurate where they finish, but might fail to complete many segments: ORB-SLAM3 and cuVSLAM each complete 19 of 35 segments and none on Oxford Spires, whereas \ourmethod completes all of them.

Among streaming methods, \ourmethod has the lowest SE(3) APE on four of the five datasets, with the widest margins on the large-scale walking datasets GrandTour and Oxford Spires. It does poorly on OpenLORIS, which is not used for training: there, HorizonStream has a lower SE(3) APE, and LingBot-Map, ABot, and HorizonStream a lower Sim(3) APE. As in driving, the IMU leaves the translational RPE unchanged and lowers the rotational RPE on every dataset, giving a small but consistent reduction in APE.

In the batch setting, as in driving, the optimization in VGGT-SLAM2 without loop closures brings no significant gain over \ourmethodvo with the same frontend: the two are close in SE(3) APE on every dataset except Oxford Spires. Furthermore, \ourmethod has a lower SE(3) APE  on Oxford Spires, ScaRF, and OpenLORIS, and a higher one on GrandTour and ADT. While MASt3R-Fusion has the lowest SE(3) APE on GrandTour, Oxford Spires, and ScaRF, it only  completes 10 of 16 ADT segments and 1 of 5 OpenLORIS segments. Ultimately, the batched cross-attention version of \ourmethod gives no consistent gain over streaming. 

\subsection{Runtime Performance}

% Runtime scaling
%\begin{wraptable}{r}{0.65\columnwidth}
\begin{table}

\caption{\textbf{Runtime scaling.} Processing time and cumulative FPS on RTX 6000 Ada. KITTI-360: 2013-05-28, 0006, camera 00.
%Excludes io, warmup, and initialization.
}
\vspace{-5pt}

\label{tab:runtime:scaling}
\vspace{-6pt}
\centering
\begingroup
\providecommand{\tabheader}[1]{\textbf{#1}}
\providecommand{\ourmethod}{AnyVIO}
\setlength{\tabcolsep}{2pt}
\renewcommand{\arraystretch}{1.05}
\footnotesize
%\resizebox{0.65\columnwidth}{!}{%
\begin{NiceTabular}{c|cc|cc|cc}[
    code-before =
    \rectanglecolor{gray!12}{3-4}{5-5}
    \rectanglecolor{gray!25}{3-6}{5-7}
]
\toprule
 & \multicolumn{2}{c}{\tabheader{FoundationVIO}
 %\shortstack{FoundationVIO\\{\scriptsize (corrected)}}
 } & \multicolumn{2}{c}{
 \tabheader{\ourmethod{}}
 %\shortstack{\textbf{\ourmethod{}}\\{\scriptsize w/o CG}}
 } & \multicolumn{2}{c}{
 \tabheader{\ourmethod{}$^\dagger$}
 %\shortstack{\textbf{\ourmethod{} (CG)}\\{\scriptsize w/ CG}}
 } \\
\tabheader{Frames} & \tabheader{Time (s)} & \tabheader{FPS} & \tabheader{Time (s)} & \tabheader{FPS} & \tabheader{Time (s)} & \tabheader{FPS} \\
\midrule
316 & 39.15 & 8.07 & 27.90 & 11.33 & 17.67 & 17.88 \\
1,999 & 587.32 & 3.40 & 176.24 & 11.34 & 113.35 & 17.64 \\
4,798 & 5\,480.39 & 0.88 & 420.91 & 11.40 & 273.66 & 17.53 \\
\bottomrule
\end{NiceTabular}
%}
\par\vspace{2pt}
{\raggedright\footnotesize
$\dagger$: Cuda graph optimization enabled.
%Continuous prefixes; one pass. Excludes loading, warmup/capture, VI initialization and final optimization. Eager/graph poses match exactly. \ourmethod{}: 4 CPU threads; FoundationVIO: native frontend, saved CPU settings.
\par}
\endgroup
%\end{wraptable}
\vspace{-12pt}

\end{table}

\begin{wraptable}{r}{0.5\columnwidth}
\vspace{-12pt}

\caption{\textbf{Throughout and delay} on
RTX 6000 Ada.
\tabheader{FPS}: frames per second (Hz); \tabheader{Delay}: mean image-to-pose latency (ms).}

\label{tab:runtime:methods}
\vspace{-9pt}
\centering
\def\runtimeShade#1#2{\colorbox{#1}{#2}}

\resizebox{0.5\columnwidth}{!}{\begin{NiceTabular}{cl|cc}[
            code-before =%
            % VO best
            \rectanglecolor{\secondbestcolor}{2-3}{2-4}
            \rectanglecolor{\bestcolor}{3-3}{3-4}
            % VIO best
            \rectanglecolor{\bestcolor}{10-3}{10-4}
            \rectanglecolor{\secondbestcolor}{12-3}{12-4}
            % VO best
            \rectanglecolor{\secondbestcolor}{16-3}{16-4}
            \rectanglecolor{\bestcolor}{15-3}{15-4}
            % VIO best
            \rectanglecolor{\secondbestcolor}{17-3}{17-4}
            \rectanglecolor{\bestcolor}{19-3}{19-4}
            ]
\toprule
\multicolumn{2}{c|}{\tabheader{Method}} &
\tabheader{FPS} &
\tabheader{Delay} \\
\midrule

\multirow{11}{*}{\rotatebox[origin=c]{90}{{\color{gray}Incremental / streaming}}} &\diminish{ORB-SLAM3} &  \diminish{37.2} & \diminish{28} \\
&\diminish{cuVSLAM} &  \diminish{80.2} & \diminish{15} \\
&HorizonStream &  12.4 & 82 \\
&LingBot-Map &  14.7 & 62 \\
&STream3R &  13.1 & 78 \\
&ABot & 12.5 & 78 \\
&\textbf{\ourmethodvo{} (stream)}   & 
13.1 &
76 \\
&\textbf{\ourmethodvo{}$^{\dagger}$ (stream)} &
22.8 &
42 \\
\cmidrule{2-4}
&BotVIO & 112.5 & 9 \\
&\textbf{\ourmethod{} (stream)}  &
9.8 &
237 \\
&\textbf{\ourmethod{}$^{\dagger}$ (stream)}  &
14.5 &
70 \\
\midrule

\multirow{7}{*}{\rotatebox[origin=c]{90}{{\color{gray}Batch / submap}}} &VGGT-SLAM2 & 12.3 & $2044$ \\
&MapAnything  & 23.2 & $1348$ \\
&VGGT-$\Omega$  & 34.2 & $1141$ \\
&\textbf{\ourmethodvo{} (batch)} & 
31.9 &
$1172$ \\
\cmidrule{2-4}
&MASt3R-Fusion & 11.0 & 1980 \\
&FoundationVIO  & 8.5 & $4079$ \\
&\textbf{\ourmethod{} (batch)}  &
18.9 &
1482 \\
\bottomrule
\end{NiceTabular}}
\par\vspace{2pt}
{\raggedright\footnotesize
$\dagger$: Cuda graph optimization enabled.
\par}
\vspace{-12pt}

\end{wraptable}
\ourmethod fuses images and inertial measurements in a single forward pass with bounded memory and compute. We demonstrate this by measuring \emph{throughout} as the number of frames processed per second of wall time (FPS), excluding I/O, model loading, and model warm-up, and \emph{pose delay} as the average time from when an image is acquired to when the system publishes its camera pose. \cref{tab:runtime:scaling} shows this on a KITTI-360 drive. \ourmethod's throughput stays constant, both without and with CUDA graphs~\cite{gray2019cudagraphs}. FoundationVIO, whose factor graph grows with the sequence length, falls from 8.1 to 0.9\,FPS and needs 91 minutes for this 8-minute recording, 20 times longer than AnyVIO$^\dagger$. In~\cref{tab:runtime:methods}, \ourmethod has a throughput of 14.5\,FPS and a delay of 70\,ms, only 28\,ms later than \ourmethodvo. While BotVIO is the fastest VIO method, its translational RPE on driving is above 65\% (\cref{tab:main:rpe}). \ourmethod is metric, accurate, within a real-time budget, and with no optimizer to tune.

\begin{table}[t!]
\centering
\caption{\textbf{Effect of inertial fusion.}
Equal-weight dataset averages: 11 datasets for APE,
nine with eligible 100\,m intervals for RPE.}
\vspace{-10pt}

\label{tab:ablation:imu}
\resizebox{0.7\columnwidth}{!}{\begin{NiceTabular}{l|rrrr}[
            code-before =%
            % VO best
            \rectanglecolor{\secondbestcolor}{4-2}{4-5}
            \rectanglecolor{\bestcolor}{5-2}{5-5}
            ]
\toprule
 & \multicolumn{2}{c}{\tabheader{APE (m)}}
 & \multicolumn{2}{c}{\tabheader{RPE / 100\,m}} \\
\tabheader{Variant} & \tabheader{SE(3)} & \tabheader{Sim(3)} & \tabheader{Trans. (\%)} & \tabheader{Rot. ($^\circ$)} \\
\midrule
Fine-tuned VO & 11.19 & 8.13 & 5.48 & 6.77 \\
+ Fusion Head w/o IMU & 11.12 & 7.93 & 5.40 & 6.32 \\
+ IMU fusion & {5.86} & {3.48}
             & {4.55} & {2.61} \\
\bottomrule

\end{NiceTabular}}
\vspace{-12pt}
\end{table}

\subsection{Ablations}
In Tab.~\ref{tab:ablation:imu}, \ourmethod's fusion head performs better with IMU inputs than without them and than the vision-only variant. Additional method ablations are in the supplementaries.

\subsection{Hardware results}\label{sec:hwresults}
Next, we demonstrate \ourmethod on real hardware in the \texttt{tarmac} and \texttt{offroad} racing settings of the \ourdataset (\cref{fig:mosaic}). 
%The first is a 2-mile-long \texttt{tarmac} setting with few visual features and significant glare (see Fig~.\ref{fig:mosaic}). The car is driven at speeds exceeding 40\,m/s with moderate IMU noise. The second is an \texttt{offroad} setting, where a different car is driven aggressively on a 1.6-mile-long course through clouds of dust with sideslip in excess of 50$^{\circ}$ and vertical accelerometer noise is an order of magnitude larger due to chassis vibrations (see Fig~.\ref{fig:mosaic}).
The \texttt{tarmac} vehicle runs inference on an onboard 4070 Ti S GPU, and the \texttt{offroad} vehicle uses an onboard L4 GPU. Both cars use a version of \ourmethod with a causal visual frontend fine-tuned on a subset of \ourdataset without Cuda graph optimization. %It is causal and  publishes poses in blocks of 16 frames rather than after every frame. 
The results in~\cref{tab:hardware:oncar} are grouped in nominal and out-of-distribution deployments, and the numbers are computed in real time on the vehicles. Additional experiments and details are in the supplementaries, and the video attachment includes live visualizations.
%are in~\cref{app:harware:additional}.

\noindent\textbf{Nominal Conditions.} In these experiments, the scene is visually similar to the training data, and both lighting and visibility are good. %The cars are driven counterclockwise, despite a vast majority of the training data for \texttt{offroad} being clockwise. 
The ordering is as expected: \ourmethod is more accurate than \ourmethodvo, which is far more accurate than the IMU model alone (SE(3) APE above 100\,m). On \texttt{tarmac}, the IMU lowers the translational RPE from 6.57\% to 4.35\% and the SE(3) APE from 29.98 to 17.03\,m. On \texttt{offroad}, \ourmethod reaches an SE(3) APE of 3.76\,m.

\noindent\textbf{Out-of-Distribution Robustness Tests.} %The previous experiments were all in a different season from the bulk of the fine-tuning data, but t
To control for overfitting and analyze robustness, we conduct a series of experiments in settings that are OOD by design: ``\texttt{tarmac}, dark'' is tested in poor lighting that differs significantly from training; ``\texttt{tarmac}, reverse'' involves driving the track in the reverse (CW) direction with good visibility; and ``\texttt{offroad}, dusty'' has poor visibility and significant glare.

\begin{figure}[t!]
    \centering
    \begin{tikzpicture}
        \node[anchor=south east, draw=black, inner sep=0pt, name=A] at (-0.2cm,0.05cm) {\adjincludegraphics[trim={{.05\width} {.0\height} {.05\width} 0}, clip, height=2cm]{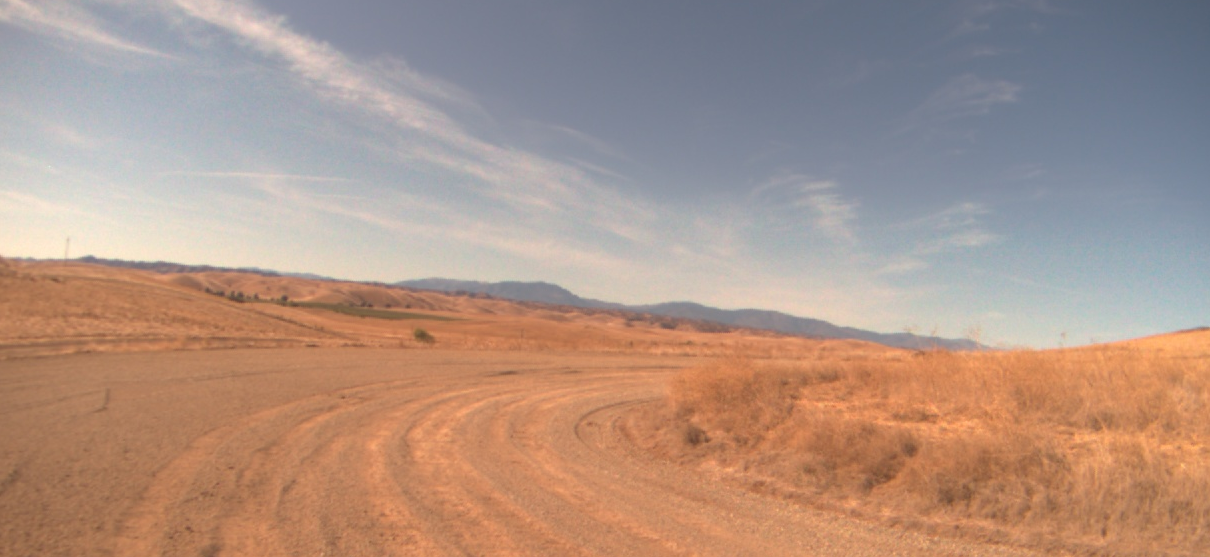}};
        \node[anchor=north east, draw=black, inner sep=0pt, name=B] at (-0.2cm,-0.05cm) {\adjincludegraphics[trim={{.05\width} {.0\height} {.05\width} 0}, clip, height=2cm]{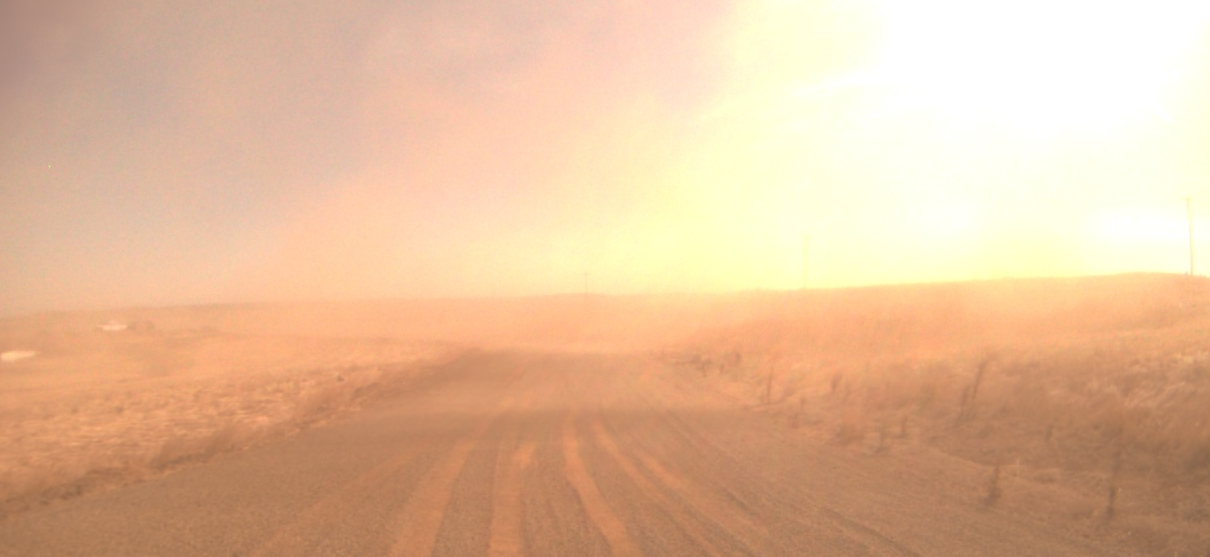}};
        \node[anchor=west] at (0,0) {\includegraphics[height=5cm]{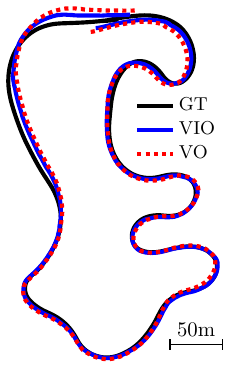}};
        \draw[black, dotted] (A.east) -- (2.5cm,0.0cm);
        \draw[black, dotted] (B.east) -- (1.0cm,-0.5cm);
    \end{tikzpicture}

    \vspace{-10pt}

    \caption{\textbf{Out-of-distribution results from on-board inference.} SE(3) aligned trajectories in the dusty \texttt{offroad} experiment.}
    \label{fig:OODtrajectories}
    \vspace{-12pt}
\end{figure}

In poor lighting (see~\cref{fig:mosaic}), the odometry errors match those in broad daylight. In the reverse direction, the IMU helps less: the SE(3) APE of \ourmethod increases to 26.71\,m, close to that of \ourmethodvo (27.52\,m), although the RPE remains lower. In heavy dust, \ourmethod keeps an SE(3) APE of 6.72\,m, against 10.53\,m without the IMU, and an RPE below that of daylight track driving (\cref{tab:hardware:oncar,fig:OODtrajectories}).

\begin{table}[t!]
    \caption{\textbf{On-car results.} RPE per 100\,m (translation in \%, rotation in $^\circ$) and APE (m). $v$: max speed (m/s). $\beta$: max sideslip. }
    \label{tab:hardware:oncar}
    \vspace{-7pt}
    \centering
    \setlength{\tabcolsep}{2.5pt}
    \resizebox{\columnwidth}{!}{%
    \begin{NiceTabular}{@{}lcc|cccc|cccc@{}}\toprule
         & & & \multicolumn{4}{c|}{\tabheader{\ourmethodvo}} & \multicolumn{4}{c}{\tabheader{\ourmethod}} \\
         & & & \multicolumn{2}{c}{RPE} & \multicolumn{2}{c|}{APE} & \multicolumn{2}{c}{RPE} & \multicolumn{2}{c}{APE} \\
         \tabheader{Condition} & $v$ & $\beta$ & Tr. & Rot. & SE(3) & Sim(3) & Tr. & Rot. & SE(3) & Sim(3) \\ \midrule
         \texttt{tarmac}$^\S$, nominal & 38 & 10$^{\circ}$ & 6.57 & 1.43 & 29.98 & 17.82 & 4.35 & 1.14 & 17.03 & 11.37 \\
         \texttt{offroad}, nominal & 23 & 38$^{\circ}$ & 3.57 & 1.54 & 6.17 & 5.60 & 2.77 & 0.85 & 3.76 & 3.06 \\ \midrule
         \texttt{tarmac}, dark$^\S$ & 39 & 5$^{\circ}$ & 5.55 & 1.39 & 27.97 & 16.05 & 3.78 & 1.15 & 15.88 & 12.15 \\
         \texttt{tarmac}, reverse & 32 & 4$^{\circ}$ & 6.08 & 1.51 & 27.52 & 18.56 & 4.90 & 1.21 & 26.71 & 16.62 \\
         \texttt{offroad}, dusty & 26 & 54$^{\circ}$ & 4.12 & 1.68 & 10.53 & 10.38 & 3.22 & 1.04 & 6.72 & 6.06 \\ \bottomrule
    \end{NiceTabular}}
    \par\vspace{2pt}
{\raggedright\footnotesize
$\S$: mean of two runs. Sideslip is defined as $\beta = \arctan(v_y^B/v_x^B)$ with velocities expressed in a vehicle-fixed frame. See e.g.,~\cite{djeumou2024one}.
\par}
\vspace{-10pt}
\end{table}

\iffalse
%\subsection{Ablations}

\mg{
Here are some suggestions. I think most of these should be done, and most will probably end up in the appendix.

\noindent\textbf{Visual backbone.}

\begin{itemize}
    \item Submap overlap in VO: Defining $a_k$ such that we have one overlapping frame (default) all the way to N-1 overlapping frames (low-latency streaming). 
    \item Ablate on the GRU architecture in the scale head. This seems different from the MA scale head. Why did we depart? What motivated the use of GRU+MLP here?
    \item Ablate on the impact of the confidence statistics in~\eqref{eq:visualfeatures}. Are they necessary? What happens if you omit the mask here? Is the mask head necessary?
\end{itemize}
}

\oa{\begin{itemize}
    \item In respect to the mask head that was added purely for auxiliary purposes, and for TH specific scene reconstruction, so I do not think I need to expand on it too much.
    \item i can do the confidence stats + rescaling with the fusion head
    \item i think original bidirection and causal is covered in the main table now
\end{itemize}
}

\TODO{List all intended ablations, add preliminary results.}

\fi

%\newpage
\section{Conclusion}

%To the best of our knowledge, 
We propose \ourmethod, the first feed-forward visual-inertial streaming method, which combines a foundation model for 3D reconstruction and a foundation model for inertial sensing.
\ourmethod adapts a non-metric 3D foundation model to causal, incremental inference with bounded memory, and learned metric scale. The two motion estimates are fused according to their predicted reliability, so the cost per frame is constant and every pose is published 70\,ms after receiving the corresponding image. Across driving, indoor, and walking datasets, the inertial model improves accuracy over the visual model alone. These results support our hypothesis that foundation models for vision and inertial sensing can be combined without a geometric backend to yield robust metric odometry in challenging real-world environments.

\textbf{Limitations.} \ourmethod has no loop closure or global refinement, and batch methods therefore remain more accurate on several datasets. While the model runs on datasets not seen in training without modification, its metric scale is less accurate there than on the training domains. The cost of the inertial model also depends on the IMU stream, which can lower the frame rate. Finally, similar to other VIOs, \ourmethod requires the camera-IMU extrinsic calibration. % between the camera and the IMU, and in our deployment tests an inaccurate calibration degraded its accuracy.

\section*{Acknowledgement}
The Authors acknowledge John Subosits for helpful feedback on the manuscript and its presentation. We also thank Geoff Budd, Phung Ngyen, Michael Thompson, Steven Goldine, and Kazunori Nimura for thir support and work on the hardware platforms used in data collection and experimentation.

\section*{LLM Usage Statement}
Claude is used for code prototyping, refactoring, and running experiments. ChatGPT is used to fix syntax and grammar errors of the main text. 

{
    \small
    \bibliographystyle{ieeenat_fullname}
    \bibliography{main}
}

\end{document}